\documentclass[twocolumn]{aastex62}
\usepackage{xspace}
\usepackage{multirow}

\newcommand{\ntarg}{26\xspace}
\newcommand{\ntoi}{31\xspace}

\newcommand{\TESS}{{\em TESS}}

\newcommand{\referee}{} 

\shorttitle{OrCAS I. Survey Definition}
\shortauthors{Crossfield, et al.}

\begin{document}

\title{OrCAS: Origins, Compositions, and Atmospheres of Sub-neptunes. I.  Survey Definition}

\correspondingauthor{Ian J.\ M.\ Crossfield}
\email{ianc@ku.edu}

\author{Ian J.\ M.\ Crossfield}
\affiliation{Department of Physics and Astronomy, University of Kansas, Lawrence, KS, USA}
\affiliation{Max-Planck-Institut f\"ur Astronomie, K\"onigstuhl 17, D-69117 Heidelberg, Germany}

\author{Alex S.\ Polanski}
\affiliation{Department of Physics and Astronomy, University of Kansas, Lawrence, KS, USA}
\affiliation{Lowell Observatory, 1400 W. Mars Hill Rd., Flagstaff, AZ, 86001, USA}

\author{Paul Robertson}
\affiliation{Department of Physics and Astronomy, University of California, Irvine, CA 92697, USA}

\author{Joseph Akana Murphy}
\affiliation{Department of Astronomy and Astrophysics, University of California, Santa Cruz, CA 95060, USA}

\author{Emma V.\ Turtelboom}
\affiliation{Department of Astronomy, University of California Berkeley, Berkeley, CA 94720, USA}

\author[0000-0002-4671-2957]{Rafael Luque}
\affiliation{Department of Astronomy \& Astrophysics, University of Chicago, Chicago, IL 60637, USA}
\affiliation{NHFP Sagan Fellow}

\author{Thomas Beatty}
\affiliation{Department of Astronomy, University of Wisconsin-Madison, Madison, WI 53706, USA}

\author{Tansu Daylan}
\affiliation{Department of Physics and McDonnell Center for the Space Sciences, Washington University, St. Louis, MO 63130, USA}

\author[0000-0002-0531-1073]{Howard Isaacson}
\affiliation{Department of Astronomy, University of California Berkeley, Berkeley, CA 94720, USA}

\author{Jonathan Brande}
\affiliation{Department of Physics and Astronomy, University of Kansas, Lawrence, KS, USA}

\author{Laura Kreidberg}
\affiliation{Max-Planck-Institut f\"ur Astronomie, K\"onigstuhl 17, D-69117 Heidelberg, Germany}

\author{Natalie M.\ Batalha}
\affiliation{Department of Astronomy and Astrophysics, University of California, Santa Cruz, CA 95060, USA}

\author{Daniel Huber}
\affiliation{Institute for Astronomy, University of Hawai'i, 2680 Woodlawn Drive, Honolulu, HI 96822, USA; Sydney Institute for Astronomy (SIfA), School of Physics, University of Sydney, NSW 2006, Australia}

\author{Maleah Rhem}
\affiliation{Department of Physics and Astronomy, University of Kansas, Lawrence, KS, USA}

\author{Courtney Dressing}
\affiliation{Department of Astronomy, University of California Berkeley, Berkeley, CA 94720, USA}

\author{Stephen R.\ Kane}
\affiliation{Department of Earth and Planetary Sciences, University of California, Riverside, CA 92521, USA}

\author{Malik Bossett}
\affiliation{Department of Astronomy and Astrophysics, University of California, Santa Cruz, CA 95060, USA}

\author{Anna Gagnebin}
\affiliation{Department of Astronomy and Astrophysics, University of California, Santa Cruz, CA 95060, USA}

\author[0009-0002-2757-4138]{Maxwell A.\ Kroft}
\affiliation{Department of Astronomy, University of Wisconsin-Madison, Madison, WI 53706, USA}

\author[0000-0001-5728-4735]{Pranav H.\ Premnath}
\affiliation{Department of Physics and Astronomy, University of California, Irvine, CA 92697, USA}

\author[0009-0005-5520-1648]{Claire J.\ Rogers}
\affiliation{Department of Physics and Astronomy, University of California, Irvine, CA 92697, USA}


\author[0000-0001-6588-9574]{Karen A.\ Collins}
\affiliation{Center for Astrophysics \textbar \ Harvard \& Smithsonian, 60 Garden Street, Cambridge, MA 02138, USA}

\author[0000-0001-9911-7388]{David W. Latham}
\affiliation{Center for Astrophysics \textbar \ Harvard \& Smithsonian, 60 Garden Street, Cambridge, MA 02138, USA}

\author[0000-0001-8621-6731]{Cristilyn N.\ Watkins}
\affiliation{Center for Astrophysics \textbar \ Harvard \& Smithsonian, 60 Garden Street, Cambridge, MA 02138, USA}

\author[0000-0002-5741-3047]{David R.\ Ciardi} 
\affiliation{NASA Exoplanet Science Institute, IPAC, California Institute of Technology, Pasadena, CA 91125 USA}

\author[0000-0002-2532-2853]{Steve B.\ Howell} 
\affiliation{NASA Ames Research Center, Moffett Field, CA 94035, USA}

\author[0000-0002-2454-768X]{Arjun B. Savel} 
\affiliation{Department of Astronomy, University of Maryland, College Park, College Park, MD 20742 USA}

\author{Perry Berlind} 
\affiliation{Center for Astrophysics \textbar \ Harvard \& Smithsonian, 60 Garden Street, Cambridge, MA 02138, USA}

\author{Michael L. Calkins} 
\affiliation{Center for Astrophysics \textbar \ Harvard \& Smithsonian, 60 Garden Street, Cambridge, MA 02138, USA}

\author{Gilbert A. Esquerdo} 
\affiliation{Center for Astrophysics \textbar \ Harvard \& Smithsonian, 60 Garden Street, Cambridge, MA 02138, USA}

\author{Jessica Mink} 
\affiliation{Center for Astrophysics \textbar \ Harvard \& Smithsonian, 60 Garden Street, Cambridge, MA 02138, USA}

\author[0000-0002-2361-5812]{Catherine A. Clark} 
\affiliation{Jet Propulsion Laboratory, California Institute of Technology, Pasadena, CA 91109 USA}
\affiliation{NASA Exoplanet Science Institute, IPAC, California Institute of Technology, Pasadena, CA 91125 USA}

\author[0000-0003-2527-1598]{Michael B. Lund} 
\affiliation{NASA Exoplanet Science Institute, IPAC, California Institute of Technology, Pasadena, CA 91125 USA}

\author[0000-0001-7233-7508]{Rachel~A.~Matson} 
\affiliation{U.S. Naval Observatory, Washington, D.C. 20392, USA}

\author[0000-0002-0885-7215]{Mark~E.~Everett}  
\affiliation{U.S. National Science Foundation National Optical-Infrared Astronomy Research Laboratory, 950 N.\ Cherry Ave., Tucson, AZ 85719, USA}

\author{Joshua E. Schlieder}  
\affiliation{NASA Goddard Space Flight Center, Exoplanets and Stellar Astrophysics Laboratory, Mail Code 667, 8800 Greenbelt Rd, Greenbelt, MD 20771, USA}

\author[0000-0003-0593-1560]{Elisabeth C.\ Matthews}  
\affiliation{Max-Planck-Institut f\"ur Astronomie, K\"onigstuhl 17, D-69117 Heidelberg, Germany}

\author[0000-0002-8965-3969]{Steven Giacalone}  
\altaffiliation{NSF Astronomy and Astrophysics Postdoctoral Fellow}
\affiliation{Department of Astronomy, California Institute of Technology, Pasadena, CA 91125, USA}

\author[0000-0001-7139-2724]{Thomas Barclay} 
\affiliation{NASA Goddard Space Flight Center, Exoplanets and Stellar Astrophysics Laboratory, Mail Code 667, 8800 Greenbelt Rd, Greenbelt, MD 20771, USA}

\author{Roberto Zambelli}
\affiliation{Societa Astronomica Lunae, Castelnuovo Magra, Italy}

\author{Peter Plavchan}
\affiliation{George Mason University, 4400 University Dr, Fairfax, VA 22030}

\author{Taylor Ellingson} 
\affiliation{George Mason University, 4400 University Dr, Fairfax, VA 22030}

\author{Michael Bowen}  
\affiliation{George Mason University, 4400 University Dr, Fairfax, VA 22030}

\author{Gregor Srdoc}
\affil{Kotizarovci Observatory, Sarsoni 90, 51216 Viskovo, Croatia}

\author[0000-0001-9504-1486]{Kim K. McLeod}
\affiliation{Department of Physics and Astronomy, Wellesley College, Wellesley, MA 02481, USA}

\author[0000-0001-8227-1020]{Richard P. Schwarz}
\affiliation{Center for Astrophysics \textbar \ Harvard \& Smithsonian, 60 Garden Street, Cambridge, MA 02138, USA}

\author[0000-0003-1464-9276]{Khalid Barkaoui}
\affiliation{Astrobiology Research Unit, Universit\'e de Li\`ege, 19C All\'ee du 6 Ao\^ut, 4000 Li\`ege, Belgium}
\affiliation{Department of Earth, Atmospheric and Planetary Science, Massachusetts Institute of Technology, 77 Massachusetts Avenue, Cambridge, MA 02139, USA}
\affiliation{Instituto de Astrof\'isica de Canarias (IAC), Calle V\'ia L\'actea s/n, 38200, La Laguna, Tenerife, Spain}

\author{Jacob Kamler}
\affiliation{John F. Kennedy High School, 3000 Bellmore Avenue, Bellmore, NY 11710, USA}

\author[0000-0001-9087-1245]{Felipe Murgas}
\affiliation{Instituto de Astrof\'isica de Canarias (IAC), Calle V\'ia L\'actea s/n, 38200, La Laguna, Tenerife, Spain}
\affiliation{Departamento de Astrof\'isica, Universidad de La Laguna (ULL), E-38206 La Laguna, Tenerife, Spain}

\author{Enric Palle}
\affiliation{Instituto de Astrof\'isica de Canarias (IAC), Calle V\'ia L\'actea s/n, 38200, La Laguna, Tenerife, Spain}
\affiliation{Departamento de Astrof\'\i sica, Universidad de La Laguna (ULL), 38206, La Laguna, Tenerife, Spain}

\author[0000-0001-8511-2981]{Norio Narita}
\affiliation{Komaba Institute for Science, The University of Tokyo, 3-8-1 Komaba, Meguro, Tokyo 153-8902, Japan}
\affiliation{Astrobiology Center, 2-21-1 Osawa, Mitaka, Tokyo 181-8588, Japan}
\affiliation{Instituto de Astrof\'isica de Canarias (IAC), Calle V\'ia L\'actea s/n, 38200, La Laguna, Tenerife, Spain}

\author[0000-0002-4909-5763]{Akihiko Fukui}
\affiliation{Komaba Institute for Science, The University of Tokyo, 3-8-1 Komaba, Meguro, Tokyo 153-8902, Japan}
\affiliation{Instituto de Astrof\'isica de Canarias (IAC), Calle V\'ia L\'actea s/n, 38200, La Laguna, Tenerife, Spain}

\author[0009-0009-5132-9520]{Howard M. Relles}
\affiliation{Center for Astrophysics \textbar \ Harvard \& Smithsonian, 60 Garden Street, Cambridge, MA 02138, USA}

\author[0000-0001-6637-5401]{Allyson Bieryla}
\affiliation{Center for Astrophysics \textbar \ Harvard \& Smithsonian, 60 Garden Street, Cambridge, MA 02138, USA}

\author{Eric Girardin}
\affiliation{Grand Pra Observatory, 1984 Les Hauderes, Switzerland}

\author[0000-0001-8879-7138]{Bob Massey}
\affil{Villa '39 Observatory, Landers, CA 92285, USA}

\author[0000-0003-2163-1437]{Chris Stockdale}
\affiliation{Hazelwood Observatory, Australia}

\author[0000-0003-0828-6368]{Pablo Lewin}
\affiliation{The Maury Lewin Astronomical Observatory, Glendora,California.91741. USA}

\author[0009-0006-9361-9153]{Riccardo Papini}
\affiliation{Wild Boar Remote Observatory, San Casciano in val di Pesa, Firenze, 50026 Italy}

\author[0000-0002-0619-7639]{Pere Guerra}
\affiliation{Observatori Astron\`omic Albanyà, Cam\'i de Bassegoda S/N, Albany\`o 17733, Girona, Spain}

\author[0000-0003-2239-0567]{Dennis M.\ Conti}
\affiliation{American Association of Variable Star Observers, 185 Alewife Brook Parkway, Suite 410, Cambridge, MA 02138, USA}

\author{Sel\c{c}uk Yal\c{c}inkaya}
\affiliation{Ankara University, Faculty of Science, Astronomy and Space Sciences Department, Tandogan, TR-06100, Ankara, T\"urkiye}
\affiliation{Ankara University, Astronomy and Space Sciences Research and Application Center (Kreiken Observatory), Alper Gezeravc{\i} Yerle\c{s}kesi, {\.I}ncek Blvd., TR-06837, Ahlatlıbel, Ankara, T\"urkiye}

\author{\"Ozg\"ur Ba\c{s}t\"urk}
\affiliation{Ankara University, Faculty of Science, Astronomy and Space Sciences Department, Tandogan, TR-06100, Ankara, T\"urkiye}
\affiliation{Ankara University, Astronomy and Space Sciences Research and Application Center (Kreiken Observatory), Alper Gezeravc{\i} Yerle\c{s}kesi, {\.I}ncek Blvd., TR-06837, Ahlatlıbel, Ankara, T\"urkiye}

\author{Ghachoui Mourad}
\affiliation{Oukaimeden Observatory, High Energy Physics and Astrophysics Laboratory, Cadi Ayyad University, Marrakech, Morocco}

\begin{abstract}
Sub-Neptunes --- volatile-rich exoplanets smaller than Neptune --- are
intrinsically the most common type of planet known.  However, the
formation and nature of these objects, as well as the distinctions
between sub-classes (if any), remain unclear.  Two powerful tools to
tease out the secrets of these worlds are measurements of (i)
atmospheric composition and structure revealed by transit and/or
eclipse spectroscopy, and (ii) mass, radius, and density revealed by
transit photometry and Doppler spectroscopy.  Here we present OrCAS, a
survey to better elucidate the origins, compositions, and atmospheres
of sub-Neptunes.  This radial velocity survey uses a repeatable,
quantifiable metric to select targets suitable for subsequent
transmission spectroscopy and address key science themes about the
atmospheric \& internal compositions and architectures of these
systems.  Our survey targets \ntarg\ systems with transiting
sub-Neptune planet candidates, with the overarching goal of increasing
the sample of such planets suitable for subsequent atmospheric
characterization.  This paper lays out our survey's science goals,
defines our target prioritization metric, and performs light-curve
fits and statistical validation using existing \TESS\ photometry and
ground-based follow-up observations.  Our survey serves to continue
expanding the sample of small exoplanets with well-measured properties
orbiting nearby bright stars, ensuring fruitful studies of these
systems for many years to come.

\end{abstract}


\section{Introduction}
\label{sec:intro}

Precise elemental and molecular abundance measurements provide some of the strongest
constraints on the formation and evolution of planets, stars, and
galaxies.  In the coming decades, such measurements will also inform the search for
extrasolar biosignatures.   JWST is now the superlative platform
for  exoplanet characterization because it allows us to finally
measure precise atmospheric abundances
\citep[e.g.,][]{ers:2023,madhusudhan:2023b,bell:2023,beatty:2024}.
To provide a clearer understanding of planet formation, composition,
and chemistry, we must build a complete, homogeneous sample of
planetary abundances across wide swaths of parameter space and link
together the compositions of exoplanets, their host stars, and the
Solar system.

\begin{figure*}[]
\includegraphics[width=\textwidth]{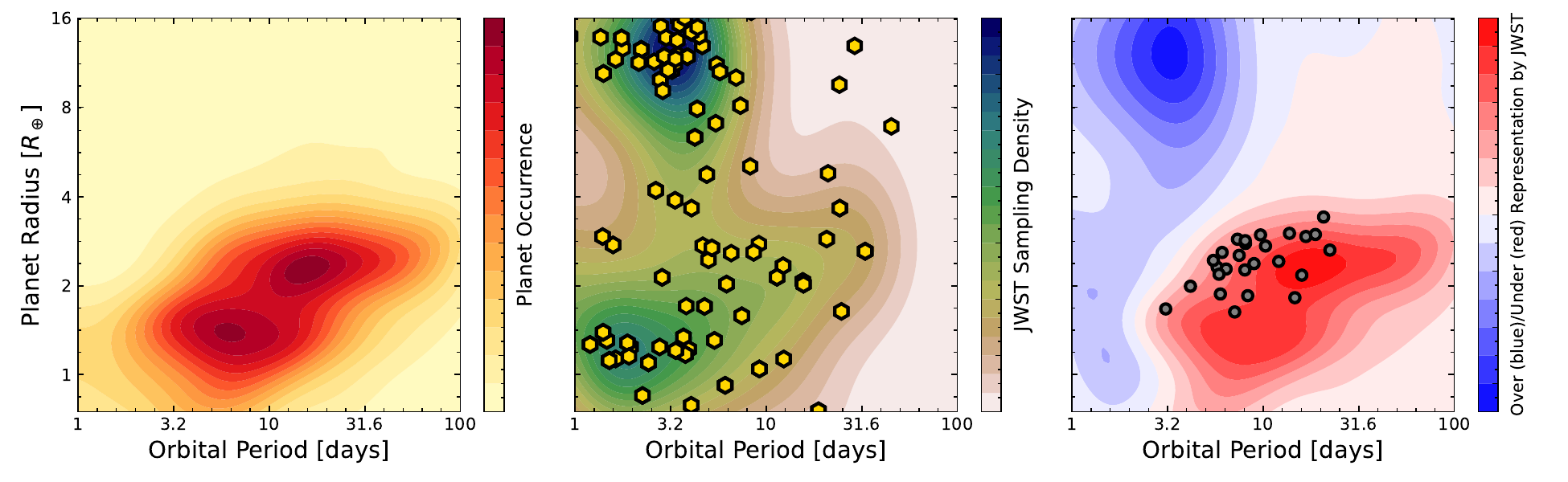}
\caption{ Target prioritization. {\em Left:} Planet occurrence rate
  \citep{fulton:2018}, showing that sup-Neptunes and super-Earths are
  the most common planets on short-period orbits. {\em Center:} Gold
  hexagons show transiting exoplanets targeted by JWST in Cycles
  1 \& 2 (our survey began before Cycle 3 results were announced); the shading is a Kernel Density Estimate map showing that hot
  Jupiters and super-Earths were the most commonly targeted
  planets. {\em Right:} Difference of the two heat maps, showing that
  temperate sub-Neptunes (red region) remain underrepresented by JWST
  relative to the intrinsic underlying population. We identified our
  targets (gray points) by weighting by this under-representation, TSM,
  and observing efficiency (see text for details).  
\label{fig:triptych}}
\end{figure*}

Sub-Neptunes and super-Earths occur more frequently than any other
type of known exoplanet, making these bodies the most common end
product of the process of planet formation
\citep[Fig.~\ref{fig:triptych}a;][]{howard:2010,fulton:2017}. {\referee
  Although in general sub-Neptunes are larger and more volatile-rich
  than super-Earths, it is challenging to discriminate the two classes
  based on transit-derived properties alone. Thus although our work
  targets the more H/He-rich sub-Neptunes, some smaller, denser
  super-Earths may also make it into our sample.  For either of these
  planet types, observing the planets'} atmospheres provides one of
the best hopes for determining formation history, and the {\em only}
hope for understanding the interplay between initial composition and
present-day chemistry, dynamics \& circulation, and disequilibrium
processes \citep[][]{bowler:2016,deming:2017}.  Models of planet
formation for sub-Neptunes can be directly tested via JWST
measurements of atmospheric composition, but to make the greatest
impact precise ($\ge$5$\sigma$) masses are needed to prevent mass
uncertainties from dominating the error budgets of atmospheric
retrievals \citep{batalha:2019}.

Furthermore, precise planet masses are a looming, underappreciated
bottleneck for JWST.  According to the NASA Exoplanet
Archive\footnote{\url{https://exoplanetarchive.ipac.caltech.edu/}},
over two thousand sub-Neptunes are known but only 167 have 5$\sigma$
masses; of those, only 81 are good targets for atmospheric
characterization (TSM\footnote{\citep[Transmission Spectroscopy
  Metric, a proxy for the S/N of JWST transmission
  spectroscopy;][]{kempton:2018}.}$>$30).  JWST (and soon, ARIEL) need
hundreds more such planets if they are to successfully survey large
numbers of exoplanet atmospheres.  Confirming these planets via
precise RV masses, and preparing for atmospheric spectroscopy, is the
primary (though not sole) focus of our program.

This program is OrCAS: an RV survey to inform our understanding of the
Origins, Compositions, and Atmospheres of Sub-neptune exoplanets.  By
using modern RV facilities such as WIYN/NEID \citep{schwab:2016b} and
Keck/KPF \citep{gibson:2016} the program aims to provide a large
number (up to \ntarg) of sub-Neptunes with precise masses, in order to
support subsequent JWST and ARIEL atmospheric characterization of a
large and demographically-representative sample of these
intrinsically-common planets. In addition, our program will
characterize stellar activity in known planet-hosting systems, explore
system architectures, and determine precise planet densities and bulk
compositions. OrCAS  currently obtains its observations with Keck/KPF
\citep{gibson:2016} through a four-semester NASA Keck Strategic
Mission Support program (KSMS), and with WIYN/NEID
\citep{schwab:2016b} through allocations from NOIRLab and the
University of Wisconsin-Madison (UW-Madison).

Given the heady success of TESS at finding new planets that are
subjected to subsequent characterization, naturally ours is far from
the only effort focusing on sub-Neptune characterization. A
non-exhaustive list of these includes the PFS Magellan-TESS Survey
\citep{teske:2021}, the HIRES TESS-Keck Survey
\citep{chontos:2022,polanski:2024}, the HARPS NCORES survey
(Armstrong et al., in review), and more.

This paper defines the OrCAS survey, and is written in the same spirit
as \cite{teske:2021} and \cite{chontos:2022}; our first results and
mass measurements, for TOI-6054/HD~23074, are presented in Paper~II
(Kroft et al., in prep.). Here we present our science goals,
rigorously describes our approach to target selection, and provides
transit fits and statistical validation of our target sample.

\section{Science Themes}
\label{sec:themes}
\begin{figure}
\centering
\includegraphics[width=0.5\textwidth]{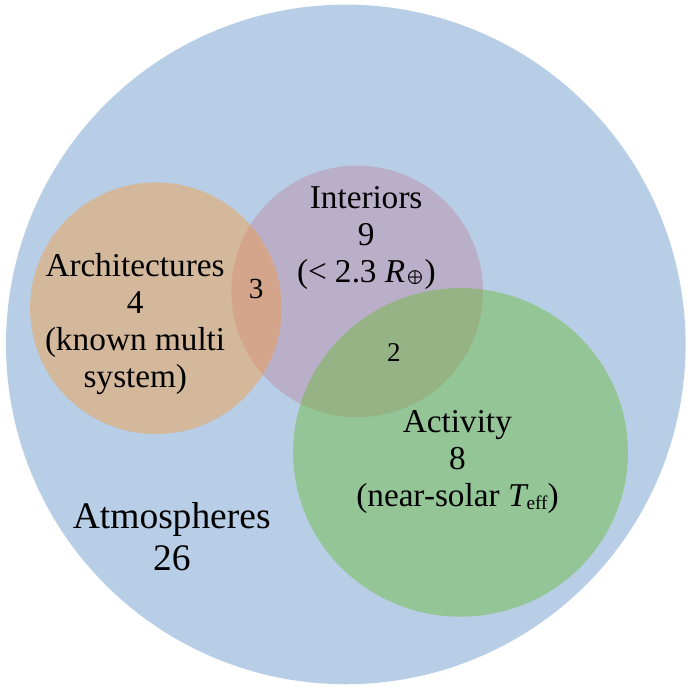}
\caption{Number of target systems in each of our science themes:
  atmospheres (I), interior compositions (II), system architectures
  (III), and stellar activity (IV). See Sec.~\ref{sec:themes} and Table~\ref{tab:systems} for details.}
\label{fig:venn}
\end{figure}

Here we present the four main science themes of our program. The
primary and driving goal (Theme I, Sec.~\ref{sec:projectone} below)
is to enable many more future sub-Neptune atmospheric measurements by
obtaining $\ge5\sigma$ RV planet masses.  Table~\ref{tab:systems}
and Fig.~\ref{fig:venn} show the breakdown of which targets fall under
which science theme.

\begin{deluxetable*}{l r l l l r l l c l r}
\tabletypesize{\scriptsize}
\tablecaption{  Stellar Parameters (from TIC): \label{tab:systems}}
\tablewidth{0pt}
\tablehead{
\colhead{TOI} & \colhead{TIC} & \colhead{Names} & \colhead{RA [$^o$]} & \colhead{Dec [$^o$]} & \colhead{$V$ [mag]} & \colhead{$T_\mathrm{eff}$ [K]} & \colhead{$R_*\ [R_\odot]$} & \colhead{$M_*\ [M_\odot]$} &  \colhead{Theme} & \colhead{$n_\mathrm{sectors}$} 
}
\startdata
1174 & 154089169 & TYC 4403-2080-1 & 209.217175 & +68.618211 & 10.96 & 5030$\pm$ 60 & 0.760$\pm$0.050 & 0.85$\pm$0.10  & 1  & 10  \\ 
1180 & 158002130 & TYC 4565-656-1 & 214.555774 & +82.193764 & 11.02 & 4738$\pm$139 & 0.730$\pm$0.050 & 0.75$\pm$0.09  &  1 & 14  \\ 
1184 & 233087860 & TYC 4201-1875-1 & 272.20452 & +60.678785 & 10.99 & 4534$\pm$128 & 0.690$\pm$0.060 & 0.71$\pm$0.08  & 1  & 31  \\ 
1244 & 219850915 & TYC 4421-473-1 & 256.279559 & +69.519093 & 11.93 & 4599$\pm$123 & 0.721$\pm$0.053 & 0.73$\pm$0.09  & 1  &  29 \\ 
1247 & 232540264 & HD135694,\,HIP74326 & 227.867864 & +71.841296 & 9.08 & 5712$\pm$106 & 1.080$\pm$0.050 & 1.02$\pm$0.12  & 1\,\, \,\ 4& 16  \\ 
1301 & 356867115 & TYC 4437-668-1 & 275.862543 & +71.581748 & 11.34 & 4820$\pm$ 32 & 0.732$\pm$0.048 & 0.77$\pm$0.09  & 1\,\ 3  &  30 \\ 
1467 & 240968774 & LP 151-22,\,NLTT4212 & 19.114644 & +49.233138 & 12.29 & 3834$\pm$157 & 0.491$\pm$0.015 & 0.489$\pm$0.020  & 1\,2  &  3 \\ 
1630 & 317597583 & BD+74 1020 & 352.016676 & +75.552607 & 9.62 & 5504$\pm$130 & 0.830$\pm$0.040 & 0.95$\pm$0.13  &  1 & 7  \\ 
1659 & 288631580 & BD+82 553 & 277.925194 & +82.305949 & 10.08 & 5977$\pm$131 & 1.006$\pm$0.048 & 1.10$\pm$0.14  &  1\,2\,3 & 16  \\ 
1691 & 268334473 & BD+86 279 & 272.404118 & +86.860343 & 10.13 & 5759$\pm$187 & 1.000$\pm$0.050 & 1.03$\pm$0.14  & 1\,\ \,\ 4  & 11  \\ 
1716 & 14336130 & HD237566 & 105.083329 & +56.824194 & 9.41 & 5879$\pm$141 & 1.240$\pm$0.070 & 1.07$\pm$0.14  & 1\,\ \,\ 4  & 4  \\ 
1723 & 71431780 & BD+68 495 & 116.797198 & +68.476685 & 9.66 & 5777$\pm$125 & 1.090$\pm$0.050 & 1.04$\pm$0.13  &  1\,\ \,\ 4 & 5  \\ 
1739 & 159418353 & LP 8-221,\,NLTT42078 & 240.172481 & +83.258659 & 10.69 & 4814$\pm$125 & 0.781$\pm$0.054 & 0.78$\pm$0.09  & 1\,2  & 15  \\ 
1744 & 229455001 & BD+60 1982 & 294.291071 & +60.881302 & 9.54 & 6179$\pm$125 & 1.140$\pm$0.050 & 1.18$\pm$0.16  &  1 &  30 \\ 
1758 & 367858035 & TYC 4602-357-1 & 354.741581 & +75.685232 & 10.79 & 5169$\pm$122 & 0.810$\pm$0.050 & 0.88$\pm$0.11  & 1  & 7  \\ 
1768 & 137906197 & TYC 3818-1387-1 & 152.102783 & +55.53996 & 10.42 & 5994$\pm$121 & 1.100$\pm$0.050 & 1.10$\pm$0.15  & 1  & 4  \\ 
1772 & 85293053 & BD+35 2144 & 156.837704 & +34.39095 & 9.96 & 5583$\pm$110 & 0.995$\pm$0.046 & 0.99$\pm$0.13  &  1\ \,3 &  2 \\ 
1777 & 29191624 & HD87620,\,HIP49576 & 151.807032 & +46.114881 & 9.31 & 5859$\pm$102 & 0.980$\pm$0.040 & 1.06$\pm$0.13  & 1\,2\ 4  &  2 \\ 
1782 & 160045097 & UCAC4 823-018203  & 213.078772 & +74.411791 & 12.33 & 4336$\pm$121 & 0.669$\pm$0.056 & 0.68$\pm$0.08  & 1\,2\,3  & 11  \\ 
1798 & 198153540 &  TYC 3465-1049-1 & 211.093937 & +46.519283 & 11.36 & 5165$\pm$104 & 0.780$\pm$0.040 & 0.87$\pm$0.11  & 1\,2\,3  &  6 \\ 
1799 & 8967242 & HD96735,\,HIP54491 & 167.232915 & +34.303015 & 8.98 & 5690$\pm$123 & 0.957$\pm$0.046 & 1.01$\pm$0.13  & 1\,2\,\ 4  & 2  \\ 
2211 & 212253390 & HD195755,\,HIP101503 & 308.539086 & $-$29.672613 & 9.99 & 5253$\pm$123 & 0.834$\pm$0.047 & 0.90$\pm$0.11  &  1\,2 & 2  \\ 
4638 & 58824746 & TYC 611-390-1 & 12.820663 & +12.787986 & 11.66 & 5700$\pm$131 & 0.895$\pm$0.044 & 1.02$\pm$0.13  & 1\,\ \,\ 4  &  3 \\ 
5726 & 71582156 &  BD+69 458& 123.39505 & +68.57459 & 9.74 & 5874$\pm$119 & 1.106$\pm$0.051 & 1.06$\pm$0.13  & 1\,\ \,\ 4  & 5 \\ 
5799 & 328081248 & NLTT48731,\,G 143-29 & 301.630179 & +15.989142 & 13.29 & 3521$\pm$157 & 0.343$\pm$0.010 & 0.327$\pm$0.020  & 1\,2  & 1 \\ 
6054 & 392681545 & HD23074,\,HIP17540  & 56.337007 & +60.350917 & 8.02 & 6112$\pm$119 & 1.612$\pm$0.070 &  1.15$\pm$0.16  &  1\ \,3 & 1 \\
\enddata
\end{deluxetable*}

\subsection{Theme I: Atmospheres}
\label{sec:projectone}
Volatile elements such as C and O have long been observed in exoplanet
atmospheres \citep[e.g.,][]{konopacky:2013}, and refractory elements
have also been recently measured in hot Jupiters
\citep[e.g.,][]{lothringer:2021,tsai:2023,sing:2024,welbanks:2024}. JWST
is finally beginning to allow us to extend these compositional
analyses to reveal the makeup and formation pathways of sub-Neptunes
\citep{madhusudhan:2023b,crossfield:2023,beatty:2024}.

Fundamentally, measuring atmospheric compositions of sub-Neptunes
requires atmospheric spectroscopy -- and thus also requires precise
masses ($\le$20\% fractional precision) to enable that spectroscopy
\citep{batalha:2019}. With JWST's unprecedented sensitivity,
constraints on atmospheric composition are finally precise enough to
begin to test the quantitative predictions of planet formation models.

For example, early JWST transmission spectroscopy of K2-18b highlights
the sub-Neptune atmospheric science now enabled by JWST
\citep{madhusudhan:2023b}.  These spectra revealed CH$_4$ and CO$_2$
and show that JWST is capable of efficiently producing high-precision
measurements even for small planets.  The measured abundances for the
identified species are far more precise than was feasible with HST,
with JWST breaking key model degeneracies (e.g., between H$_2$O and
CH$_4$). Such tight constraints on atmospheric composition would not
be possible without the 5.4$\sigma$ planet mass constraint obtained
from RV observations \citep{cloutier:2019,radica:2022}.

However, JWST is rapidly running out of small transiting planet
targets. In its first three Cycles, JWST will characterize 133
transiting exoplanets, from hot Jupiters to cool rocky worlds
(Fig.~\ref{fig:triptych}b). If JWST operates for 20 years then the
mission could characterize the atmospheres of nearly a thousand
exoplanets during its lifetime. Such a large survey may be necessary
to beat down the stochasticity of planet formation and finally enable
 population-level studies linking atmospheres to planet size,
mass, orbits, and host star properties. To best
investigate the smaller sub-Neptunes that are the most common
end-products of planet formation, most of JWST's effort should focus
on these intrinsically common worlds. However, sub-Neptunes are
extremely underrepresented in JWST's target pool, especially compared
to intrinsic occurrence rates (Fig.~\ref{fig:triptych}).

Furthermore, JWST's accessible, small-planet sample could be exhausted
in just $\sim$two more cycles \citep[leaving none for
  ARIEL;][]{tinetti:2018} unless additional sub-Neptune masses can be
measured. Though there are currently 94 known planets with $R_p < 4
R_\oplus$, high TSM, and a $>$5$\sigma$ mass, JWST is observing 56
planets with $R_p < 4 R_\oplus$ in Cycles~1--3. As described below,
our new TESS-discovered systems will be excellent targets for future
GO programs to explore the atmospheres of sub-Neptune exoplanets.

Completing a demographic assay of sub-Neptune atmospheres will take
years and require a much larger sample of sub-Neptunes with precise
masses. Efforts to date (mainly with HST) have been suggestive of
trends of cloud properties with equilibrium temperature
\citep[e.g.,][]{brande:2024}, but it remains unclear whether these
conclusions are consistent with recent JWST results \citep[e.g.,][and
  others]{wallack:2024,alderson:2024,beatty:2024}.  A larger number of
deep, sensitive, JWST observations are urgently needed.  Above all
else, our program's measurement of these new sub-Neptune masses from
1.6--4.2\,$R_\oplus$ and $T_\mathrm{eq}$ from 400--1200~K
(Fig.~\ref{fig:sample}) aims to support the burgeoning field of
sub-Neptune atmospheric characterization.

\subsection{Theme II: Internal Compositions}
Radius alone is not sufficient to infer even an approximate internal
composition for sub-Neptunes, since the bulk makeup of these planets
becomes highly degenerate between rock, ice, and H$_2$/He
\citep{adams:2008,figueira:2009}. Furthermore, masses of the
sub-Neptune population can vary by up to a factor of 5 for a given
radius implying a compositional diversity not reflected by
single-valued mass-radius relations \citep{wolfgang:2016,otegi:2020,parviainen:2024}.  Systems
such as TOI-849 \citep[3.5 $R_\oplus$ with a mass of 39
  $M_\oplus$;][]{armstrong:2020} and TOI-1853 \citep[3.5 $R_\oplus$,
  but a whopping 73 $M_\oplus$;][]{naponiello:2023} demonstrate that
in the sub-Neptune population, both radius and mass are needed to
reliably infer density and surface gravity, and to begin to infer bulk
planetary composition.

With precise masses from Doppler spectroscopy and refined radii from
transit analyses, derived measurements of bulk density can contribute
to population-level searches for demographic trends. Such measurements
also serve to constrain planets' core mass fractions and elemental
compositions, and provide hints about how and where they formed
\citep{luque:2022}.  For example, rocky and giant planets are expected
to form within and beyond the frost line, respectively, whereas water
worlds may have formed outside the frost line but have migrated inward
since formation.  TESS has observed numerous transits of most of our
sample (see Table~\ref{tab:systems}), enabling precise radius
measurements of our planet candidates (see Sec.~\ref{sec:transits}).

Recent RV and transit surveys suggest that below
$\lesssim$2$R_\oplus$, the super-Earth/sub-Neptune dichotomy may
actually be a trichotomy of rocky, icy, and gaseous small planets
orbiting cooler M dwarfs \citep{luque:2022}. Probing the nuances of
this region of the mass-radius diagram has largely been inhibited by a
lack of precise mass measurements (especially for planets around
hotter, more massive stars).  More data points are needed to explore
this part of mass-radius space in finer detail {\referee \citep{parc:2024}}, and to determine
whether the compositional model of \cite{luque:2022} extends to
planets orbiting stars hotter than M dwarfs {\referee \citep{rogers:2023}}. 

More broadly, there is considerable interest in measuring the shape
and intrinsic scatter of the exoplanetary mass-radius relation.  By
providing an objective, reproducible target prioritization metric (see
Sec.~\ref{sec:selection}) we aim to allow our mass and radius
measurements to be incorporated into future efforts, e.g. using
hierarchical Bayesian modeling and non-parametric mass-radius relations
\citep{wolfgang:2016,ning:2018,kanodia:2019,kanodia:2023}, with minimal selection bias effects.

\subsection{Theme III: System Architectures}

Sub-Neptunes are observed to be the most commonly-occurring product of planet formation
\citep[e.g.,][]{fulton:2018}; their orbital architectures and atmospheres
 are complementary fossils of their formation and
migration histories.  For example, when planets smoothly migrate through
their natal disk the systems may remain flat, densely-packed, and
dynamically `cool,' while systems that evolved by planet-planet
scattering may become more widely spaced and dynamically `hot.'  Measuring the
planet multiplicity, orbital eccentricity, and spin-orbit alignment
(obliquity) constrains these pathways \citep{winn:2015}.

Measuring the masses of multiple planets in a single system enables
comparative atmospheric characterization, e.g.\ by controlling for the
unknown X-ray and UV irradiation history as well as our imprecise
knowledge of stellar age and abundances \citep{owen:2019,libby:2020}.
By providing precise mass measurements of all known planets in
multi-planetary systems, we aim to deliver cosmic laboratories
amenable to atmospheric studies of multi-planet systems.

Modern RV follow-up of TESS planets also serves to bring Kepler-era
results into the TESS era --- for example, testing for ``Peas in a
Pod'' \citep[e.g.,][]{weiss:2018} by comparing orbits, sizes, masses,
and eccentricities of our sample; using long-baseline RVs to test the
``Edge-of-the-multis'' claim that multi-planet systems appear
truncated relative to detection limits \citep{millholland:2022}; and
allowing an independent, RV-driven test of the eccentricity peak seen
at 2\,$R_\oplus$ in Kepler systems \citep[][]{gilbert:2024}.  Our long
RV baseline ($\gtrsim$2~yrs) and high precision will likely reveal
additional non-transiting and/or long-period companions, whose
discovery will improve the interpretation of these systems.  Such
long-term RV studies reveal, e.g., the connection between the
formation of short-period sub-Neptunes and cold gas giant planets on
orbits more like those found in our own solar system
\citep[e.g.,][]{bonomo:2023,bitsch:2023}. Precise masses and Keplerian orbits will
improve dynamical stability modeling of our targets, providing
evidence for companions or resonances, as well as determining regions
where dynamically stable orbits are excluded.

Finally, precise RVs can also measure the sky-projected obliquity
between a planet's orbital and its star's rotational axes via the
Rossiter-McLaughlin (RM) effect. A large spin-orbit misalignment may
be indicative of secular gravitational interactions with an outer
companion \citep{winn:2010,bourrier:2021}. Whether by measuring our
targets' masses or (in later programs) measuring orbital obliquities,
our validated sample will enable the confirmation of these candidates
as true planets and facilitate later atmospheric studies of these worlds.

\subsection{Theme IV: Stellar Activity}
\label{sec:theme4}
With next-generation Doppler spectrometers, the chief RV limitation is
often astrophysical noise arising from variable stellar surfaces.
Developing improved techniques for diagnosing and mitigating this
stellar ``jitter'' is a critical requirement for the extreme-precision
radial velocity (EPRV) community to achieve its near- and long-term
goals \citep{crass:2021} of measuring masses of transiting planets for
atmospheric study \citep{batalha:2019}, and identifying targets for
NASA's Habitable Worlds Observatory \citep[HWO;][]{luhn:2023}.
High-precision RV surveys provide legacy data sets that are immensely
valuable for testing strategies for identifying and removing RV jitter
around a broad range of stellar types using cutting-edge data.

Several aspects of our experimental design will make the resulting
time-series data ideal for developing techniques to mitigate
astrophysical noise.  Our targets are all transiting systems, which
``freezes'' the planetary periods and ephemerides and simplifies the
challenge of modeling planets and jitter simultaneously.  Furthermore,
the TESS discovery light curves provide additional information on the
variability driving jitter, since even non-contemporaneous space-based
light curves help predict and correct for RV jitter
\citep{aigrain:2012,bastien:2013,beard:2024b}.  The broad wavebands,
intrinsic stability, and high resolution, of modern spectrographs
allow for the extraction of a diverse set of stellar activity tracers,
such as absorption-line indicators (Ca II H\&K, H$\alpha$, Ca IR
triplet), line-shape deformation metrics \citep{zechmeister:2018}, and
chromatic dependence of signal amplitudes \citep{cale:2021}.  The high
S/N achieved is especially important for ``line-by-line'' RV
extractions, which minimize jitter by excluding activity-sensitive
absorption lines \citep[e.g.][]{cretignier:2023,ould-elhkim:2023}.

We expect that several of our targets will ultimately require
intensive activity modeling to obtain robust planetary masses.  This
activity challenge is an opportunity to use precise RV data to refine
jitter mitigation techniques and push the limiting exoplanet detection
sensitivity down to the level of intrinsic instrumental noise
floors. This work will support future EPRV efforts to seek all kinds
of planets -- including small, rocky, temperate HWO targets orbiting
Sun-like stars -- especially since our target $T_\mathrm{eff}$
distribution closely mirrors the HWO target list \citep{mamajek:2024}.

\section{Target Selection, Vetting, and Validation}
\label{sec:selection}

\subsection{Target Selection}
We designed our program to have a well-defined, repeatable target
selection process in order to facilitate subsequent, minimally-biased
population-level studies.  The primary consideration in designing the
sample and associated prioritization metrics was the goal of expanding
the pool of sub-Neptunes with $\ge$5$\sigma$ masses so that future
JWST and ARIEL atmospheric spectroscopy can focus on a planet sample
more generally representative of underlying demographic trends
(cf.\ Fig.~\ref{fig:triptych}).

To prioritize our targets, we considered all TOIs as of September 2023
and followed the steps below. First, we took the intrinsic
short-period planet occurrence distribution \citep{fulton:2018}. Since
the underlying period-radius distribution from that work was not
available from the authors (B.\ Fulton, priv.\ comm.)\ we constructed
our own proxy for the intrinsic occurrence rate as follows.  We took
the merged list of physical parameters from the California-Kepler
Survey \citep{petigura:2017,johnson:2017} then followed an analogous
set of steps as that described by \cite{fulton:2018}: specifically,
selecting only those systems with impact parameter $b<0.7$, period $P
< 100$~d, stellar radii $R_* < 2.1 R_\odot$, stellar $T_\mathrm{eff}$
between 4700--6500~K, and finite planet radii. This results in the
period-size distribution shown in Fig.~\ref{fig:triptych}a, which
closely resembles the corresponding distribution of
\cite{fulton:2018}.

We then constructed a Kernel Density Estimate (KDE) map from the
approved JWST Cycle 1 and 2 transiting exoplanet sample (green shading
in Fig.~\ref{fig:triptych}b). The difference between these two maps
shows which types of planets are over- or under-represented (blue and
red regions, respectively, in Fig.~\ref{fig:triptych}c) in JWST's
target list relative to intrinsic occurrence rates. We dub this the
Sub-neptune Under-representation Rate (SUR).  Each of the three color
maps in Fig.~\ref{fig:triptych} is available in electronic form, to
facilitate reproduction of our analysis.

For each potential target, we then estimated the necessary number of
RV observations (and the total observing time) to reach a 5$\sigma$
mass.  Although OrCAS radial velocity spectra are being obtained from
a variety of sources, to ensure a homogeneous, repeatable, and
straightforward target selection metric we must select a single
instrument to use for these calculations. Since our largest observing
award consists of Keck/KPF time, we used the KPF exposure time
calculator\footnote{\url{https://github.com/California-Planet-Search/KPF-etc}}
to estimate the observing time required. We assumed an internal KPF
noise floor of 0.5~m~s$^{-1}$ (KPF Instrument Team, priv.\ comm.)\ and
added estimated stellar RV noise from granulation \&\ oscillation
\citep[following][]{yu:2018rv} and rotation \citep{galland:2005}.  We
used the candidates' radii to estimate their masses and RV
semi-amplitudes via the simple, invertible, single-valued power-law
relation $M_P = {R_P}^{2.06}$ \citep[in Earth
  units;][]{lissauer:2011a}.  Then we compared the total estimated
per-shot RV uncertainty {\referee $\sigma_1$ (obtained by adding the
  expected instrumental and astrophysical noise sources in
  quadrature)} to the TOI's estimated RV semi-amplitude, {\referee $K$,
  esimating a minimum number of RVs needed to reach a 5$\sigma$ mass
  as $ \left(5 / (K / \sigma_1) \right)^2$}.

Our experience shows that for such small RV signals, and for these
smaller planets (which are frequently in multi-planet systems with
other non-transiting companions), we must obtain a minimum of
$\gtrsim$30~RVs per target \citep[see also][]{murphy:2024}. Furthermore, since programs that stop observations after
achieving a preset significance threshold (e.g., 5$\sigma$) result in
upwardly biased masses \citep{burt:2018} we therefore plan to obtain
{\referee a minimum of} 30 RVs for all targets; this choice should also
help to minimize bias in our final mass sample
\citep[cf.][]{teske:2021,chontos:2022}. {\referee Since the performance of
  KPF is still evolving, in the final survey we may elect to obtain a
  larger number of RVs for a smaller subset of targets.}

Our final prioritization metric $M$ is then the SUR
(Fig.~\ref{fig:triptych}c) of each target's radius and period,
multiplied by each target's TSM, divided by the estimated time to
measure a 5$\sigma$ mass for that planet:
\begin{equation}
  M = {\mathrm{SUR}\left( R_p, P \right)  \times \mathrm{TSM}\left( R_P, M_P, a, R_*, T_\mathrm{eff}, m_J \right) } / {{t}_{5\sigma}} .
\end{equation}
We settled on this choice in order to access not simply the best
overall targets for atmospheric study, but merely targets of
reasonably high quality whose masses were not unduly expensive to
measure.  Finally, we also required all targets to have TSM$\ge$30 and
$J>6$ mag to allow precise JWST constraints without wholly saturating
the useful instrument modes, and declination $> -20^o$ (to ensure
targets that are accessible from northern-hemisphere observatories
such as KPF and NEID). This final step left us with a list of roughly
100 rank-ordered TOIs. {\referee A Python software tool to generate our prioritized target list, along with SUR values, is available online at \url{https://github.com/iancrossfield/OrCAS_target_selection}.}

\subsection{Target Vetting and Validation}
\label{sec:validation}
To ensure that observatory resources are well-utilized, we
comprehensively vetted all of these $\sim$100 prioritized TOIs. We
used the TESS Exoplanet Follow-up Observing Program (ExoFOP) website
to vet our potential targets.  We removed all targets where:
ground-based time-series photometry (SG1, see Appendix~\ref{sec:sg1})
revealed that the transit occurs around a fainter, nearby star; where
single-epoch recon spectroscopy (SG2) revealed secondary lines, large
RV shifts indicative of stellar companions, very high activity
($R'_\mathrm{HK}>-4.5$), or $v \sin i > 7$~km~s$^{-1}$; where
high-resolution imaging (SG3, see Appendix~\ref{sec:sg3}) revealed
nearby stellar companions that could actually host the transit signal
or that would be blended with the target star in the KPF aperture;
where targets were already being intensively followed up by precise RV
campaigns (SG4); and where the TESS project's photometric vetting
plots indicated significant secondary eclipses, off-target transits,
or coherent photometric variability of $>$0.1\%. {\referee These
  checks led us to exclude the following TOIs from our sample:
  1154.01, 1262.01, 1277.02, 1287.01, 1404.01, 1430.01, 1432.01,
  1434.01, 1437.01, 1438.02, 1440.01, 1441.01, 1443.01, 1445.01,
  1451.01, 1453.01, 1464.01, 1473.01, 1643.01, 1648.01, 1730.01,
  1730.03, 1735.01, 1736.01, 1740.01, 1742.01, 1751.01, 1778.01,
  1801.01, 1802.01, 1824.01, 1835.01, 2280.01, 2287.01, 4576.01,
  5076.01, 5117.01, 5156.01, 5515.01, and 5539.01.}

We then statistically validated our remaining 29 target planet candidates
using the public \texttt{TRICERATOPS} tool \citep{giacalone:2021}, which
calculates the false positive and nearby false positive probabilities
(FPP/NFPP) of each candidate transit signal. Such validation analyses attempt
to rule out astrophysical false positive scenarios such as eclipsing
binaries and have  been employed since the
Kepler era \citep{borucki:2012}.
We begin by pulling each TOI's SPOC-processed light curve
\citep{jenkins:2016} from the MAST database using \texttt{Lightkurve}
\citep{lightkurve:2018}. The apertures used in the initial TESS alert
pipeline are also used as input to \texttt{TRICERATOPS}, as is
existing high-resolution imaging data (downloaded from the public
ExoFOP website, where available; see Appendix~\ref{sec:sg3}).  To
reliably estimate the FPP and NFPP, we run \texttt{TRICERATOPS} 20
times per candidate planet.  \cite{giacalone:2021} recommends that
after vetting, candidates be assigned a disposition of validated
planet (VP; FPP$<$0.015 and NFPP$<10^{-3}$), likely planet (LP,
FPP$<$0.5 and NFPP$<10^{-3}$), or likely nearby false positive
(NFPP$>$0.1).

Our \texttt{TRICERATOPS} analysis identified three potential targets
as having a high likelihood of being nearby false positives.  These
TOIs (and their FPP and NFPP values) are: 1432.01 (0.14$\pm$0.03 and
0.11$\pm$0.02), 1643.01 (0.32$\pm$0.15 and 0.27$\pm$0.10), and 4576.01
(0.70$\pm$0.30 and 0.70$\pm$0.30). We therefore removed these systems
from our sample, leaving our final sample of \ntarg\ targets described
in Tables~\ref{tab:systems} and~\ref{tab:params}.  The total sum of
FPP and NFPP values across our selected target sample is just 0.17,
indicating only a roughly one-in-six chance of having even a single
target turn out to be a false positive. In addition to these
\ntarg\ targeted sub-Neptunes, three systems with multiple transiting
planet signals (TOI-1782, TOI-1798, and TOI-6054/HD~23074) bring the total
number of known planets in our sample to \ntoi.

Note that although we removed potential targets being extensively
observed by high-precision Doppler spectroscopy, we elected to keep
targets being intensively observed by Keck/HIRES. This is because the
RV precision of HIRES is somewhat lower than modern RV spectrographs,
sometimes preventing it from measuring the low masses expected for
much of our sample.  Indeed, although HIRES mass measurements were
recently reported for 13 planets (including 12 in our selected
sub-Neptune sample) in Table~\ref{tab:params}, none of the selected
sub-Neptune targets rise above a significance level of 3.5$\sigma$
\citep{polanski:2024}.  The ultra-short-period planet TOI-1798.02 has
a significant mass measurement of $5.6^{+0.8}_{-0.7}\,M_\oplus$
\citep{polanski:2024}, but we include this target here only because
its companion, TOI-1798.01, is a sub-Neptune selected by our
prioritization metric.

\section{Uniform Determination of Planet Radii}
\label{sec:transits}

To derive system parameters (including transit depths and planet
radii) for our targets, we fit light curve models to the
\TESS\ photometry using the same software and approach as
\cite{crossfield:2015a,crossfield:2016}.  Although ground-based
time-series photometry exists for some targets (see
Appendix~\ref{sec:sg1}), we opt to use a homogeneous, \TESS-only
analysis to characterize our targets' transits. We used all 120-second
cadence, PDCSAP \TESS\ photometry available for each target as of May
2024, accessing the data series with \texttt{Lightkurve}
\citep{lightkurve:2018}. Table~\ref{tab:systems} lists the number of
sectors of \TESS\ data (as of mid-2024), $n_\mathrm{sectors}$, used to
analyze each system. We trimmed each light curve to include only those
points within $\pm$2.5 $T_{14}$ of the estimated transit mid-center,
reducing the data volume (and computational time) by roughly 90\%.  We
use stellar parameters and associated uncertainties from the TESS
Input Catalog (TIC) v8.2 \citep{stassun:2018b}, and used \texttt{LDTK}
\citep{parviainen:2015} to set Gaussian priors on the quadratic
limb-darkening coefficients by using the uncertainties on the stellar
parameters. {\referee Although we used the TIC to ensure homogeneous and
  repeatable target selection, subsequent analyses of individual
  systems will provide updated and refined stellar parameters, as well as bespoke transit analyses of all systems.}

\begin{figure*}
\centering
\includegraphics[width=\textwidth]{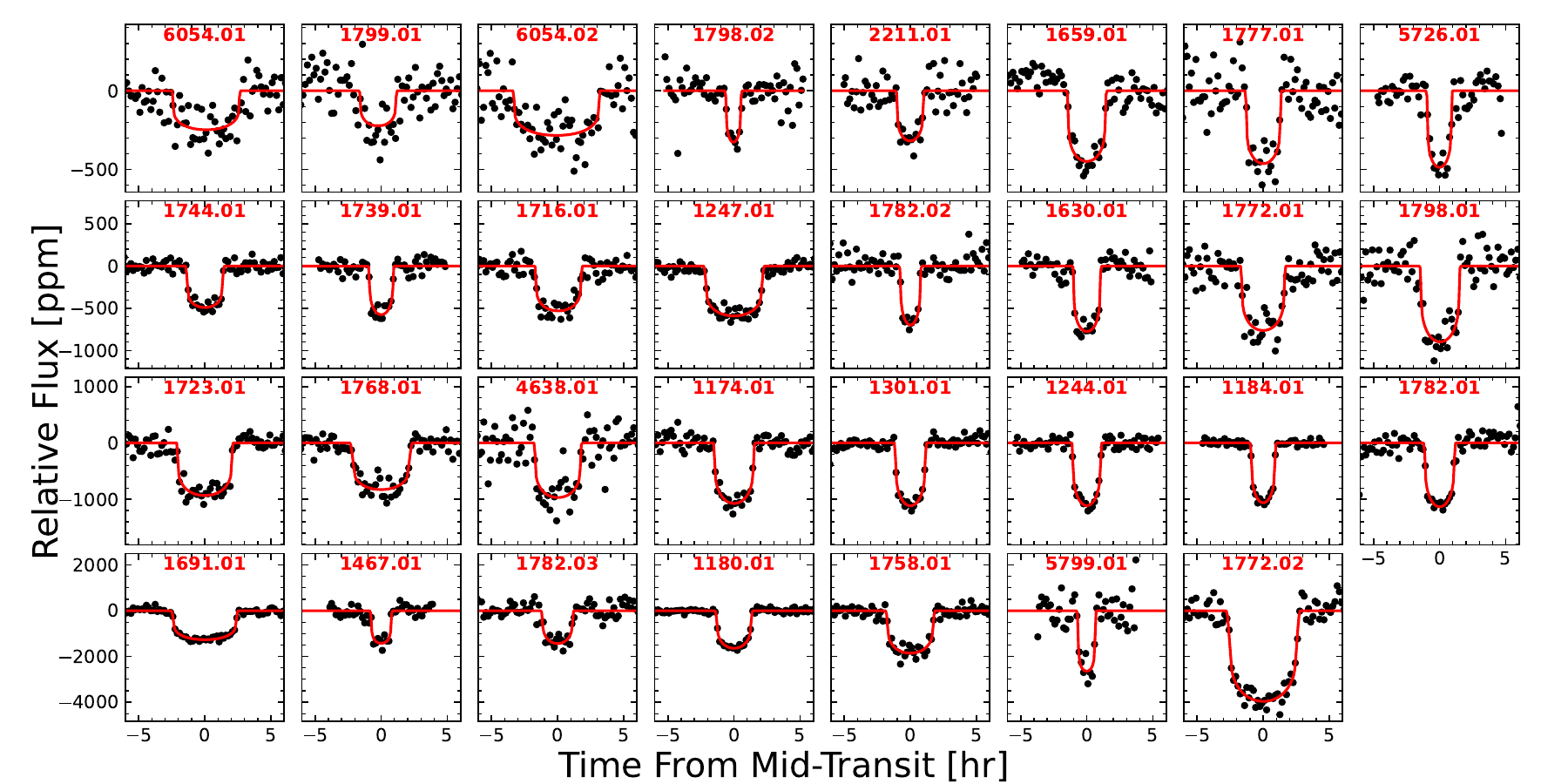}
\caption{Transit light-curve fits to our target sample, showing {\em TESS} photometry (black points, binned to a 10-minute cadence) and the best-fit light curves (red line).  }
\label{fig:transits}
\end{figure*}

Our analysis fits for transit mid-center $T_0$, orbital period $P$,
orbital inclination $i$, the ratios $R_P/R_*$ and $R_*/a$, the
quadratic limb-darkening coefficients, and an overall multiplicative
scaling constant. The light-curve analysis begins with an initial
guess defined by the TOI parameters from the TESS project, followed by
alternating applications of an optimizer \citep[the SciPy downhill simplex
algorithm, \texttt{optimize.fmin}][]{virtanen:2020} and a Markov-Chain Monte Carlo
(MCMC) sampler
\citep[\texttt{emcee};][]{foreman-mackey:2012}. \texttt{BATMAN}
\citep{kreidberg:2015b} is used to model the light curves.  The
analysis restarts when an improved fit is found, and concludes after
either 20,000 steps or when the Gelman-Rubin metric for all parameters
is $<1.03$. The final light-curve fits are shown in
Fig.~\ref{fig:transits} and the resulting parameters and uncertainties
are listed in Table~\ref{tab:params}.

\section{Target Sample \& Observing Strategy}

\subsection{Stellar \& Planetary Properties}
The stellar parameters of our final selected sample \citep[taken from
  the TESS input catalog;][]{stassun:2018b} are listed in
Table~\ref{tab:systems}. In the main our target stars are roughly
Sun-like FGK dwarfs, with only two stars hotter than 6000~K and only
two cooler than 4000~K.  The stars are moderately bright; the
brightest, TOI-6054 (HD~23074), has $m_V=8.0$~mag but the rest of the
sample has a median brightness of $m_V=10.4$~mag and $m_J=9.0$~mag
(suitable for JWST spectroscopy with most time-series modes). Stellar
activity levels, photometric variability, and rotation periods are not
known for all our stars but characterizing these properties is a key
goal of Theme~II (Sec.~\ref{sec:theme4}).

\begin{figure*}
\includegraphics[width=\textwidth]{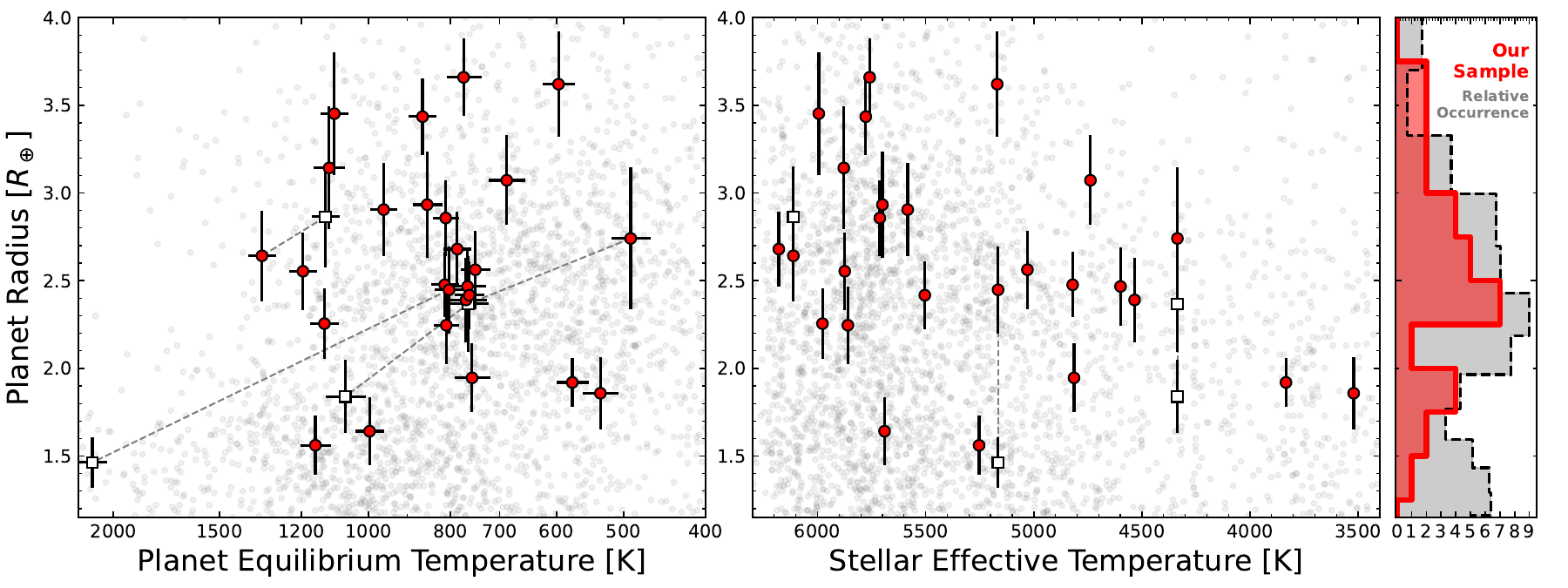}
\caption{Our target sample, showing the distribution of fit planet
  size, temperature, and stellar $T_\mathrm{eff}$. Red points are our
  primary sample, white points are known companions to our targets,
  dashed lines connect  planets orbiting the same target star, and light gray
  points are other known exoplanets. The histogram at {\em right}
  shows how our planet sizes (red) compare to the intrinsic occurrence
  of sub-Neptunes \citep[gray,][]{fulton:2018}. Noted that these
  best-fit radii differ from the TESS-reported radii plotted in
  Fig.~\ref{fig:triptych}c.
\label{fig:sample}}
\end{figure*}

Our resulting planet sample spans a wide range of sub-Neptune
parameter space.  The sample has a median radius of $2.5 R_\oplus$
with six planets larger than $3 R_\oplus$, and Fig.~\ref{fig:sample}
shows that the planets mostly lie above the radius gap
\citep{fulton:2018}. Thus the majority of our targets are likely to be
volatile-dominated water worlds or mini-Neptunes \citep{lopez:2014}.
Our targets have a median equilibrium temperature of 800~K, suggesting
that many of them may be warm enough to inhibit aerosol production and
so reveal atmospheric features in transmission \citep{brande:2024}.
The planets have a median orbital period of just 8~days; this value is
comfortably shorter than the typical rotation periods of mature
Sun-like stars, which will (we hope) result in relatively
straightforward RV analyses. Finally, our sub-Neptune targets have a
median estimated TSM of 56, indicating good prospects for atmospheric
characterization via transmission spectroscopy.  In particular,
TOI-5799.01 has an estimated TSM of 114, marking it as potentially the
best atmospheric target in our sample.  Nonetheless, given the broad
diversity of sub-Neptune compositions we expect our targets' TSM
values will change significantly from these estimated values once the
planets' masses are measured.

\subsection{Observing Strategy}
Since February 2024, OrCAS RVs are being observed with Keck/KPF
through a multi-semester NASA KSMS award and on a semester-to-semester
basis with WIYN/NEID awards through UW-Madison and NOIRLab.  Though
both instruments are both designed to achieve comparable levels of RV
precision, NEID is necessarily limited to somewhat brighter stars
given WIYN's moderate (3.5\,m) aperture diameter.  We are therefore
using NEID to observe our brightest stellar targets
(TOI-6054/HD~23074, TOI-1799/HD~96735, and TOI-1247/HD~135694; see
Table~\ref{tab:systems}) while focusing KPF on the fainter systems.  

We aim to observe all targets 1-2 times per semester with KPF to
provide sensitivity to longer-period, massive companions, and
intensively (up to $\sim$30 times) within a single observing season to
densely sample both the target planet's Doppler signal as well as
other RV signals (such as those arising from stellar activity). To the
extent possible, we attempt to adopt a similar approach with our NEID
targets. Integration times are set to achieve an internal,
instrumental Doppler precision of 0.5~m~s$^{-1}$;  for KPF this
corresponds to roughly 4--20 minutes (depending on $m_V$ and $v \sin
i$).

We aim to schedule each target's RV observations to have approximately
uniform orbital phase coverage. This approach results in the least
biased, and most significant, set of final mass measurements
\citep{burt:2018,cabona:2021}.  For the queue-based WIYN/NEID such
scheduling is relatively simple to achieve; however
classically-scheduled observations with Keck/KPF will somewhat affect
our ability to achieve an optimal observing cadence with roughly
uniform phase coverage.

\section{Conclusion}

The best hope for characterizing sub-Neptunes (the most common planets
in our Milky Way) is to undertake Doppler spectroscopy (RV) mass
measurements, followed by atmospheric characterization, of planet
candidates orbiting nearby stars. In particular, a precisely measured
mass is essential to interpret transmission spectroscopy of transiting
sub-Neptunes.  By providing a large sample of such masses, RV programs
keep the target pipeline flowing for JWST (and ARIEL) and ensure that
those missions need not `gamble' by observing targets with unknown or
only poorly-constrained masses.

Today there are over 500 sub-Neptune TOIs (planet candidates) suitable
for atmospheric characterization but lacking precise (i.e., 5$\sigma$)
masses. Since 2018 the TESS follow-up community has measured roughly
25 planet masses (or upper limits) per year. With coordination and
sufficient telescope time allocation, the community should be able to
measure the masses of all 500 sub-Neptune atmospheric targets within
the next 20 years. The OrCAS program is an integral part of
this effort.


For these reasons we have thus designed the OrCAS survey, whose
overall goals, target selection metric, and observing strategy are
presented in this paper. Our two-year mission: to seek out
newly-identified, sub-Neptune-sized \TESS\ planet candidates suitable
for atmospheric characterization via transmission spectroscopy and
then measure their masses using precise Doppler (RV) spectroscopy. We
have therefore identified a sample of \ntarg\ sub-Neptune transit
signals that are well-validated, are unlikely to be false positives, that present good
prospects for atmospheric characterization, and whose masses can be
measured with a reasonable investment of observing time. Our first
mass measurements, for the two sub-Neptunes in the TOI-6054/HD~23074 system
using WIYN/NEID radial velocities (Kroft et al., in prep.) are an
early validation of our survey strategy.  Alongside this paper, we
also present Python code to directly calculate our prioritization
metric for any given planet candidate; this should ultimately allow
target selection biases to be minimized in any subsequent analysis of
planetary mass-radius relations.

OrCAS will therefore identify high-priority targets for atmospheric
spectroscopy by measuring precise masses of new planet candidates
(Science Theme~I). Along the way we hope to also constrain interior
compositions of planets straddling the super-Earth/sub-Neptune
boundary (Theme~II), measure system architectures (Theme~III), and
quantify stellar activity in planet-hosting systems across a range of
stellar types (Theme~IV). {\referee We hope that the survey will also be
  useful for addressing additional ancillary science cases; e.g.,
  comparing stellar abundance patterns to the abundances of the stars'
  hosted planets \citep{polanski:2022,kolecki:2022,hejazi:2023}.} In
this way our survey will contribute to advancing our understanding of
the origin, compositions, and atmospheres of sub-Neptune exoplanets.


\begin{deluxetable*}{l l r l l l l l l l l}[b]
\tabletypesize{\tiny}
\tablecaption{  Transit Light Curve Parameters: \label{tab:params}}
\tablewidth{0pt}
\tablehead{
\colhead{TOI} & \colhead{$P$} & \colhead{$T_{0}$} & \colhead{$T_{14}$} & \colhead{$R_P/R_*$} & \colhead{$a$} & \colhead{$R_P$} & \colhead{$S_{inc}$} & \colhead{FPP\tablenotemark{a}} & \colhead{NFPP\tablenotemark{a}} & \colhead{Disposition\tablenotemark{b}}\\
\colhead{---} & \colhead{[d]} & \colhead{[$\mathrm{BJD_{TDB}}-2454833$]} & \colhead{[hr]} & \colhead{[\%]} & {[au]} & \colhead{[$R_\oplus$]} & \colhead{[$S_\oplus$]} & \colhead{---} & \colhead{---} & \colhead{---}
}
\startdata
1174.01 & $8.9534877^{+0.0000090}_{-0.0000103}$ & $4824.03302^{+0.00073}_{-0.00080}$ & $3.063^{+0.056}_{-0.040}$ & $3.056^{+0.249}_{-0.094}$ & $0.0799^{+0.0030}_{-0.0033}$ & $2.56^{+0.25}_{-0.20}$ & $52.0^{+9.0}_{-7.8}$             & 0.0012$\pm$0.0004 & 0$\pm$0 & VP \\              
1180.01 & $9.6867604^{+0.0000064}_{-0.0000067}$ & $5117.32735^{+0.00047}_{-0.00049}$ & $2.749^{+0.050}_{-0.035}$ & $3.85^{+0.19}_{-0.17}$ & $0.0808^{+0.0032}_{-0.0035}$ & $3.07^{+0.27}_{-0.24}$ & $37.0^{+8.1}_{-6.7}$                & 0.0035$\pm$0.0032 & 0$\pm$0 & VP\\		   
1184.01 & $5.7484338^{+0.0000026}_{-0.0000027}$ & $5127.51150^{+0.00035}_{-0.00035}$ & $1.888^{+0.033}_{-0.031}$ & $3.18^{+0.14}_{-0.17}$ & $0.0560^{+0.0022}_{-0.0023}$ & $2.39^{+0.25}_{-0.23}$ & $57^{+14}_{-11}$                    & 0.0001$\pm$0.0001 & 0$\pm$0 & VP\\		   
1244.01 & $6.4003165^{+0.0000059}_{-0.0000059}$ & $5080.80830^{+0.00073}_{-0.00070}$ & $2.210^{+0.048}_{-0.031}$ & $3.11^{+0.23}_{-0.11}$ & $0.0607^{+0.0022}_{-0.0024}$ & $2.47^{+0.24}_{-0.21}$ & $56^{+12}_{-10}$                    & 0$\pm$0 &0$\pm$0 & VP  \\			   
1247.01 & $15.923476^{+0.000020}_{-0.000020}$ & $5001.13926^{+0.00088}_{-0.00086}$ & $4.490^{+0.126}_{-0.074}$ & $2.43^{+0.14}_{-0.16}$ & $0.1247^{+0.0048}_{-0.0052}$ & $2.86^{+0.23}_{-0.21}$ & $71.8^{+11.2}_{-9.6}$                 & 0.0046$\pm$0.0133 & 0$\pm$0 & VP\\		   
1301.01 & $6.0964072^{+0.0000036}_{-0.0000036}$ & $4877.29440^{+0.00033}_{-0.00034}$ & $2.333^{+0.026}_{-0.021}$ & $3.080^{+0.139}_{-0.067}$ & $0.0599^{+0.0022}_{-0.0024}$ & $2.48^{+0.20}_{-0.18}$ & $72^{+11}_{-10}$                 & 0$\pm$0 &0$\pm$0 & VP  \\			   
1467.01 & $5.9711471^{+0.0000087}_{-0.0000085}$ & $3933.9885^{+0.0010}_{-0.0013}$ & $1.578^{+0.063}_{-0.044}$ & $3.58^{+0.30}_{-0.18}$ & $0.05075^{+0.00068}_{-0.00070}$ & $1.92^{+0.17}_{-0.11}$ & $18.1^{+3.4}_{-3.0}$                & 0.0001$\pm$0 &0.0001$\pm$0 & VP\\		   
1630.01 & $12.055735^{+0.000015}_{-0.000015}$ & $5098.22472^{+0.00091}_{-0.00110}$ & $2.087^{+0.058}_{-0.036}$ & $2.65^{+0.23}_{-0.11}$ & $0.1012^{+0.0043}_{-0.0046}$ & $2.42^{+0.23}_{-0.16}$ & $55.6^{+9.8}_{-8.3}$                  & 0.0043$\pm$0.0015 & 0.0039$\pm$0.0014  &  LP  \\ 
1659.01 & $5.877556^{+0.000015}_{-0.000018}$ & $4924.3568^{+0.0012}_{-0.0014}$ & $2.891^{+0.059}_{-0.066}$ & $2.036^{+0.229}_{-0.086}$ & $0.0658^{+0.0027}_{-0.0029}$ & $2.26^{+0.25}_{-0.16}$ & $268^{+45}_{-38}$                      & 0.0056$\pm$0.0011 & 0$\pm$0 & VP\\		   
1691.01 & $16.736889^{+0.000018}_{-0.000018}$ & $5089.72550^{+0.00073}_{-0.00071}$ & $4.980^{+0.054}_{-0.041}$ & $3.335^{+0.129}_{-0.067}$ & $0.1294^{+0.0057}_{-0.0063}$ & $3.66^{+0.24}_{-0.21}$ & $59^{+12}_{-10}$                   & 0.0001$\pm$0.0003 & 0$\pm$0 & VP\\		   
1716.01 & $8.082377^{+0.000028}_{-0.000027}$ & $5126.2213^{+0.0026}_{-0.0025}$ & $3.563^{+0.168}_{-0.100}$ & $2.32^{+0.26}_{-0.18}$ & $0.0805^{+0.0033}_{-0.0036}$ & $3.14^{+0.41}_{-0.30}$ & $254^{+47}_{-40}$                         & 0.0018$\pm$0.0002 & 0.001$\pm$0.0001&  LP  \\	   
1723.01 & $13.727013^{+0.000015}_{-0.000013}$ & $4760.95996^{+0.00071}_{-0.00081}$ & $4.233^{+0.065}_{-0.042}$ & $2.869^{+0.166}_{-0.073}$ & $0.1135^{+0.0045}_{-0.0049}$ & $3.44^{+0.24}_{-0.19}$ & $92^{+15}_{-12}$                   & 0.0006$\pm$0.0001 & 0$\pm$0 & VP\\		   
1739.01 & $8.3033414^{+0.0000092}_{-0.0000101}$ & $4798.82035^{+0.00059}_{-0.00074}$ & $1.871^{+0.047}_{-0.031}$ & $2.26^{+0.23}_{-0.10}$ & $0.0739^{+0.0029}_{-0.0031}$ & $1.95^{+0.23}_{-0.17}$ & $53.9^{+11.2}_{-9.5}$               & 0.0049$\pm$0.0009 & 0$\pm$0 & VP\\		   
1744.01 & $22.341703^{+0.000044}_{-0.000042}$ & $5075.38440^{+0.00093}_{-0.00083}$ & $2.783^{+0.065}_{-0.045}$ & $2.141^{+0.196}_{-0.090}$ & $0.1641^{+0.0073}_{-0.0080}$ & $2.68^{+0.26}_{-0.17}$ & $63.3^{+10.5}_{-8.9}$              & 0.0085$\pm$0.0024 & 0.0068$\pm$0.0021  &  LP  \\ 
1758.01 & $20.705075^{+0.000037}_{-0.000034}$ & $5091.7718^{+0.0013}_{-0.0013}$ & $3.673^{+0.080}_{-0.055}$ & $4.06^{+0.29}_{-0.14}$ & $0.1414^{+0.0054}_{-0.0059}$ & $3.62^{+0.34}_{-0.27}$ & $21.1^{+4.0}_{-3.4}$                     & 0.0002$\pm$0.0002 & 0$\pm$0 & VP\\		   
1768.01 & $7.337704^{+0.000021}_{-0.000022}$ & $4794.0224^{+0.0016}_{-0.0018}$ & $4.56^{+0.13}_{-0.11}$ & $2.87^{+0.31}_{-0.21}$ & $0.0764^{+0.0033}_{-0.0036}$ & $3.45^{+0.42}_{-0.29}$ & $240^{+39}_{-33}$                            & 0.0076$\pm$0.0035 & 0.0032$\pm$0.0007  &  LP  \\ 
1772.01 & $8.054247^{+0.000040}_{-0.000038}$ & $4795.8507^{+0.0028}_{-0.0026}$ & $3.42^{+0.13}_{-0.10}$ & $2.66^{+0.30}_{-0.14}$ & $0.0784^{+0.0032}_{-0.0035}$ & $2.91^{+0.33}_{-0.21}$ & $140^{+22}_{-19}$                            & 0.0002$\pm$0.0001 & 0.0001$\pm$0 & VP\\	   
{\referee 1772.02} & 30\tablenotemark{e} & $4785.70772^{+0.00097}_{-0.00094}$ & $5.506^{+0.109}_{-0.073}$ & $5.90^{+0.25}_{-0.13}$ & --- & $6.44^{+0.41}_{-0.34}$ & --- & --- & STPC\\
1777.01 & $14.650112^{+0.000082}_{-0.000080}$ & $4802.1067^{+0.0029}_{-0.0031}$ & $2.64^{+0.13}_{-0.12}$ & $2.09^{+0.25}_{-0.13}$ & $0.1195^{+0.0048}_{-0.0052}$ & $2.25^{+0.28}_{-0.18}$ & $71.3^{+10.6}_{-9.0}$                       & 0.067$\pm$0.0056  & 0.0005$\pm$0.0001  &  LP  \\ 
{\em 1782.01}\tablenotemark{c} & $4.9876410^{+0.0000092}_{-0.0000098}$ & $4802.7331^{+0.0011}_{-0.0010}$ & $2.328^{+0.078}_{-0.050}$ & $3.21^{+0.34}_{-0.17}$ & $0.0502^{+0.0018}_{-0.0020}$ & $2.37^{+0.31}_{-0.24}$ & $56^{+13}_{-11}$                       & 0.0028$\pm$0.0009 & 0$\pm$0 & VP \\				   
{\em 1782.02}\tablenotemark{c} & $1.8290242^{+0.0000023}_{-0.0000020}$ & $4799.81673^{+0.00056}_{-0.00061}$ & $1.544^{+0.030}_{-0.031}$ & $2.49^{+0.25}_{-0.12}$ & $0.02574^{+0.00094}_{-0.00101}$ & $1.84^{+0.23}_{-0.19}$ & $214^{+50}_{-42}$                & 0.0020$\pm$0.0002 & 0$\pm$0 & VP \\				   
1782.03 & $18.75569^{+0.00012}_{-0.00013}$ & $4792.0534^{+0.0026}_{-0.0033}$ & $2.53^{+0.26}_{-0.16}$ & $3.72^{+0.61}_{-0.33}$ & $0.1215^{+0.0044}_{-0.0048}$ & $2.74^{+0.48}_{-0.34}$ & $9.6^{+2.3}_{-1.9}$                            & 0.027$\pm$0.0041  & 0$\pm$0 & LP\\		   
1798.01 & $8.021522^{+0.000024}_{-0.000028}$ & $4855.1359^{+0.0021}_{-0.0031}$ & $3.13^{+0.17}_{-0.11}$ & $2.85^{+0.36}_{-0.15}$ & $0.0750^{+0.0030}_{-0.0032}$ & $2.45^{+0.32}_{-0.19}$ & $69.3^{+11.6}_{-9.9}$                        & 0.0003$\pm$0.0001 & 0$\pm$0 & VP\\		   
{\em 1798.02}\tablenotemark{c} & $0.43781182^{+0.00000053}_{-0.00000061}$ & $4858.31350^{+0.00071}_{-0.00082}$ & $1.215^{+0.051}_{-0.111}$ & $1.706^{+0.209}_{-0.087}$ & $0.01079^{+0.00043}_{-0.00046}$ & $1.46^{+0.18}_{-0.11}$ & $3345^{+559}_{-478}$       & 0.716$\pm$0.014 & 0.0178$\pm$0.0007 & VP\tablenotemark{d} \\
1799.01 & $7.0946^{+0.0032}_{-0.0023}$ & $4071.8176^{+0.0060}_{-0.0047}$ & $2.57^{+0.26}_{-0.17}$ & $1.57^{+0.20}_{-0.14}$ & $0.0725^{+0.0030}_{-0.0033}$ & $1.64^{+0.23}_{-0.17}$ & $164^{+27}_{-23}$                                  & 0.001$\pm$0.0001  & 0$\pm$0 & VP\\		   
2211.01 & $3.092687^{+0.000013}_{-0.000012}$ & $4205.1032^{+0.0027}_{-0.0024}$ & $2.06^{+0.12}_{-0.11}$ & $1.71^{+0.20}_{-0.12}$ & $0.0401^{+0.0016}_{-0.0017}$ & $1.56^{+0.20}_{-0.14}$ & $296^{+54}_{-46}$                            & 0.0087$\pm$0.0003 & 0.0031$\pm$0.0001  &  LP  \\ 
4638.01 & $10.283919^{+0.000093}_{-0.000099}$ & $4616.3508^{+0.0039}_{-0.0032}$ & $3.72^{+0.21}_{-0.18}$ & $2.99^{+0.37}_{-0.19}$ & $0.0930^{+0.0038}_{-0.0042}$ & $2.93^{+0.38}_{-0.24}$ & $87^{+15}_{-12}$                            & 0.0035$\pm$0.0004 & 0.0002$\pm$0 & VP \\		   
5726.01 & $5.4911252^{+0.0000095}_{-0.0000146}$ & $4010.5951^{+0.0023}_{-0.0012}$ & $1.887^{+0.048}_{-0.036}$ & $2.101^{+0.219}_{-0.095}$ & $0.0622^{+0.0025}_{-0.0027}$ & $2.55^{+0.28}_{-0.18}$ & $338^{+54}_{-46}$                   & 0.0084$\pm$0.0013 & 0.0002$\pm$0 & VP\\		   
5799.01 & $4.16466^{+0.00060}_{-0.00063}$ & $4939.3345^{+0.0020}_{-0.0020}$ & $1.51^{+0.15}_{-0.11}$ & $4.96^{+0.66}_{-0.42}$ & $0.03490^{+0.00070}_{-0.00073}$ & $1.86^{+0.26}_{-0.17}$ & $13.4^{+2.8}_{-2.4}$                         &  0.0001$\pm$0 &0.0001$\pm$0 &  VP  \\		   
6054.01 & $7.4930^{+0.0059}_{-0.0055}$ & $5080.1303^{+0.0084}_{-0.0091}$ & $5.56^{+0.41}_{-0.29}$ & $1.496^{+0.177}_{-0.092}$ & $0.0785^{+0.0034}_{-0.0038}$ & $2.64^{+0.32}_{-0.21}$ & $529^{+86}_{-72}$                               & 0.0054$\pm$0.0031 & 0$\pm$0 & VP \\
{\em 6054.02}\tablenotemark{c} & $12.5750^{+0.0070}_{-0.0072}$ & $5086.9199^{+0.0065}_{-0.0059}$ & $6.56^{+0.26}_{-0.25}$ & $1.622^{+0.202}_{-0.100}$ & $0.1109^{+0.0049}_{-0.0053}$ & $2.86^{+0.37}_{-0.23}$ & $265^{+43}_{-36}$                              & 0$\pm$0& 0$\pm$0 & VP \\
\enddata
\tablenotetext{a}{FPP and NFPP values $<10^{-4}$ are reported here as zero.}
\tablenotetext{b}{Statistical validation disposition following \cite{giacalone:2021}: VP (Validated Planet), LP (Likely Planet), {\referee or STPC (single-transit planet candidate)}.}
\tablenotetext{c}{Target names in italics are companions of our selected targets, but were not themselves selected by our prioritization metric.}
\tablenotetext{d}{Planet confirmed via RV mass measurement of $5.6^{+0.8}_{-0.7}\,M_\oplus$ \citep{polanski:2024}.}
\tablenotetext{e}{\referee Single-transit planet candidate; period set to 30~d for transit fit.}
\end{deluxetable*}


\vspace{-0.1in} \acknowledgments

We thank Dr.\ S.~Halverson for advice on the performance of both NEID and KPF, and our anonymous referee for comments that improved the quality of the final manuscript.

This research was selected under the Keck Strategic Mission Support
(KSMS) category of NASA’s Keck General Observing program which is
managed by the NASA Exoplanet Science Institute (NExScI) and funded
under the auspices of NASA’s Astrophysics Data Analysis Program (NASA
Award Number). Data presented herein were obtained at the W. M. Keck
Observatory from telescope time allocated to the National Aeronautics
and Space Administration through the agency's scientific partnership
with the California Institute of Technology and the University of
California. NASA’s time on the Keck telescopes is administered by the
NASA Exoplanet Science Institute. The Observatory was made possible by
the generous financial support of the W. M. Keck Foundation.

Funding for the TESS mission is provided by NASA’s
Science Mission directorate. We acknowledge the use of public
TESS Alert data from pipelines at the TESS Science Office and
at the TESS Science Processing Operations Center.
This research has made use of the Exoplanet Follow-up Observation
Program website, which is operated by the California Institute of
Technology, under contract with the National Aeronautics and Space
Administration under the Exoplanet Exploration Program. This paper
includes data collected by the TESS mission, which are publicly
available from the Mikulski Archive for Space Telescopes (MAST).
OrCAS is supported in part by a NASA/Keck grant 80NSSC24K1096.

The authors wish to recognize and acknowledge the very significant
cultural role and reverence that the summit of Mauna Kea has always
had within the indigenous Hawaiian community. We are most fortunate to
have the opportunity to conduct observations from this mountain.

IJMC acknowledges support from NASA ROSES grant 24-XRP24\_2-0084 , and from NASA Keck grant 80NSSC24K1096.


This work makes use of observations from the LCOGT network. Part of the LCOGT telescope time was granted by NOIRLab through the Mid-Scale Innovations Program (MSIP). MSIP is funded by NSF.


This research has made use of the Exoplanet Follow-up Observation Program (ExoFOP; DOI: 10.26134/ExoFOP5) website, which is operated by the California Institute of Technology, under contract with the National Aeronautics and Space Administration under the Exoplanet Exploration Program.


Funding for the TESS mission is provided by NASA's Science Mission Directorate. KAC and CNW acknowledge support from the TESS mission via subaward s3449 from MIT.


This is paper is based on observations made with the MuSCAT3 instrument, developed by the Astrobiology Center and under financial supports by JSPS KAKENHI (JP18H05439) and JST PRESTO (JPMJPR1775), at Faulkes Telescope North on Maui, HI, operated by the Las Cumbres Observatory.


This paper makes use of observations made with the MuSCAT2 instrument, developed by the Astrobiology Center, at TCS operated on the island of Tenerife by the IAC in the Spanish Observatorio del Teide.


Some of the observations in the paper made use of the NN-EXPLORE Exoplanet and Stellar Speckle Imager (NESSI). NESSI was funded by the NASA Exoplanet Exploration Program and the NASA Ames Research Center. NESSI was built at the Ames Research Center by Steve B. Howell, Nic Scott, Elliott P. Horch, and Emmett Quigley.

This research was carried out at the Jet Propulsion Laboratory, California Institute of Technology, under a contract with the National Aeronautics and Space Administration (80NM0018D0004).

We thank T\"UB\.{I}TAK for a partial support in using T100 telescope with project number 23BT100-2045.

The postdoctoral fellowship of KB is funded by F.R.S.-FNRS grant T.0109.20 and by the Francqui Foundation.


This work is partly supported by JSPS KAKENHI Grant Number JP24H00017,
JP24K00689, and JSPS Bilateral Program Number JPJSBP120249910.

Some of the observations in this paper made use of the High-Resolution
Imaging instrument `Alopeke and were obtained under Gemini LLP
Proposal Number: GN/S-2021A-LP-105. `Alopeke was funded by the NASA
Exoplanet Exploration Program and built at the NASA Ames Research
Center by Steve B. Howell, Nic Scott, Elliott P. Horch, and Emmett
Quigley. Alopeke was mounted on the Gemini North telescope of the
international Gemini Observatory, a program of NSF’s OIR Lab, which is
managed by the Association of Universities for Research in Astronomy
(AURA) under a cooperative agreement with the National Science
Foundation. on behalf of the Gemini partnership: the National Science
Foundation (United States), National Research Council (Canada),
Agencia Nacional de Investigacion y Desarrollo (Chile), Ministerio de
Ciencia, Tecnología e Innovacion (Argentina), Ministerio da Ciencia,
Tecnologia, Inovacoes e Comunicacoes (Brazil), and Korea Astronomy and
Space Science Institute (Republic of Korea).

Support for this work was provided by NASA through the NASA Hubble
Fellowship grant HST-HF2-51559.001-A awarded by the Space Telescope
Science Institute, which is operated by the Association of
Universities for Research in Astronomy, Inc., for NASA, under contract
NAS5-26555.

T.D. acknowledges support from the McDonnell Center for the Space Sciences at Washington University in St. Louis.


{\em Facilities:} TESS, Keck:I (KPF), WIYN (NEID), LCOGT, WIYN
(NESSI), Gemini (NIRI), Gemini (`Alopeke), Palomar (PHARO), Keck:II
(NIRC2)

{\em Software:} \texttt{emcee} \citep{foreman-mackey:2012},
\texttt{BATMAN} \citep{kreidberg:2015b}, \texttt{LDTK}
\citep{parviainen:2015}, \texttt{Lightkurve} \citep{lightkurve:2018},
\texttt{SciPy} \citep{virtanen:2020}, \texttt{matplotlib}
\citep{barrett:2005}, AstroImageJ \citep{collins:2017}, TAPIR \citep{Jensen:2013}

\bibliographystyle{apj}

\begin{thebibliography}{}
\expandafter\ifx\csname natexlab\endcsname\relax\def\natexlab#1{#1}\fi

\bibitem[{{Adams} {et~al.}(2008){Adams}, {Seager}, \&
  {Elkins-Tanton}}]{adams:2008}
{Adams}, E.~R., {Seager}, S., \& {Elkins-Tanton}, L. 2008, \apj, 673, 1160

\bibitem[{{Aigrain} {et~al.}(2012){Aigrain}, {Pont}, \&
  {Zucker}}]{aigrain:2012}
{Aigrain}, S., {Pont}, F., \& {Zucker}, S. 2012, \mnras, 419, 3147

\bibitem[{{Akana Murphy} {et~al.}(2024){Akana Murphy}, {Luque}, \&
  {Batalha}}]{murphy:2024}
{Akana Murphy}, J.~M., {Luque}, R., \& {Batalha}, N.~M. 2024, arXiv e-prints,
  arXiv:2411.02521

\bibitem[{{Alderson} {et~al.}(2024){Alderson}, {Batalha}, {Wakeford},
  {Wallack}, {Aguichine}, {Teske}, {Adams Redai}, {Alam}, {Batalha}, {Gao},
  {Kirk}, {L{\'o}pez-Morales}, {Moran}, {Scarsdale}, {Wogan}, \&
  {Wolfgang}}]{alderson:2024}
{Alderson}, L., {Batalha}, N.~E., {Wakeford}, H.~R., {et~al.} 2024, \aj, 167,
  216

\bibitem[{{Armstrong} {et~al.}(2020){Armstrong}, {Lopez}, {Adibekyan}, {Booth},
  {Bryant}, {Collins}, {Deleuil}, {Emsenhuber}, {Huang}, {King}, {Lillo-Box},
  {Lissauer}, {Matthews}, {Mousis}, {Nielsen}, {Osborn}, {Otegi}, {Santos},
  {Sousa}, {Stassun}, {Veras}, {Ziegler}, {Acton}, {Almenara}, {Anderson},
  {Barrado}, {Barros}, {Bayliss}, {Belardi}, {Bouchy}, {Brice{\~n}o}, {Brogi},
  {Brown}, {Burleigh}, {Casewell}, {Chaushev}, {Ciardi}, {Collins},
  {Col{\'o}n}, {Cooke}, {Crossfield}, {D{\'\i}az}, {Delgado Mena}, {Demangeon},
  {Dorn}, {Dumusque}, {Eigm{\"u}ller}, {Fausnaugh}, {Figueira}, {Gan},
  {Gandhi}, {Gill}, {Gonzales}, {Goad}, {G{\"u}nther}, {Helled}, {Hojjatpanah},
  {Howell}, {Jackman}, {Jenkins}, {Jenkins}, {Jensen}, {Kennedy}, {Latham},
  {Law}, {Lendl}, {Lozovsky}, {Mann}, {Moyano}, {McCormac}, {Meru},
  {Mordasini}, {Osborn}, {Pollacco}, {Queloz}, {Raynard}, {Ricker}, {Rowden},
  {Santerne}, {Schlieder}, {Seager}, {Sha}, {Tan}, {Tilbrook}, {Ting}, {Udry},
  {Vanderspek}, {Watson}, {West}, {Wilson}, {Winn}, {Wheatley}, {Villasenor},
  {Vines}, \& {Zhan}}]{armstrong:2020}
{Armstrong}, D.~J., {Lopez}, T.~A., {Adibekyan}, V., {et~al.} 2020, \nat, 583,
  39

\bibitem[{{Barrett} {et~al.}(2005){Barrett}, {Hunter}, {Miller}, {Hsu}, \&
  {Greenfield}}]{barrett:2005}
{Barrett}, P., {Hunter}, J., {Miller}, J.~T., {Hsu}, J.~C., \& {Greenfield}, P.
  2005, in Astronomical Society of the Pacific Conference Series, Vol. 347,
  Astronomical Data Analysis Software and Systems XIV, ed. P.~{Shopbell},
  M.~{Britton}, \& R.~{Ebert}, 91

\bibitem[{{Bastien} {et~al.}(2013){Bastien}, {Stassun}, {Basri}, \&
  {Pepper}}]{bastien:2013}
{Bastien}, F.~A., {Stassun}, K.~G., {Basri}, G., \& {Pepper}, J. 2013, \nat,
  500, 427

\bibitem[{{Batalha} {et~al.}(2019){Batalha}, {Lewis}, {Fortney}, {Batalha},
  {Kempton}, {Lewis}, \& {Line}}]{batalha:2019}
{Batalha}, N.~E., {Lewis}, T., {Fortney}, J.~J., {et~al.} 2019, \apjl, 885, L25

\bibitem[{{Beard} {et~al.}(2024){Beard}, {Robertson}, {Giovinazzi}, {Akana
  Murphy}, {Ford}, {Halverson}, {Han}, {Holcomb}, {Lubin}, {Luque}, {Premnath},
  {Bender}, {Blake}, {Gong}, {Isaacson}, {Kanodia}, {Li}, {Lin}, {Logsdon},
  {Lubar}, {McElwain}, {Monson}, {Ninan}, {Rajagopal}, {Roy}, {Schwab},
  {Stefansson}, {Terrien}, \& {Wright}}]{beard:2024b}
{Beard}, C., {Robertson}, P., {Giovinazzi}, M.~R., {et~al.} 2024, arXiv
  e-prints, arXiv:2408.02873

\bibitem[{{Beatty} {et~al.}(2024){Beatty}, {Welbanks}, {Schlawin}, {Bell},
  {Line}, {Murphy}, {Edelman}, {Greene}, {Fortney}, {Henry}, {Mukherjee},
  {Ohno}, {Parmentier}, {Rauscher}, {Wiser}, \& {Arnold}}]{beatty:2024}
{Beatty}, T.~G., {Welbanks}, L., {Schlawin}, E., {et~al.} 2024, \apjl, 970, L10

\bibitem[{{Bell} {et~al.}(2023){Bell}, {Welbanks}, {Schlawin}, {Line},
  {Fortney}, {Greene}, {Ohno}, {Parmentier}, {Rauscher}, {Beatty}, {Mukherjee},
  {Wiser}, {Boyer}, {Rieke}, \& {Stansberry}}]{bell:2023}
{Bell}, T.~J., {Welbanks}, L., {Schlawin}, E., {et~al.} 2023, arXiv e-prints,
  arXiv:2309.04042

\bibitem[{{Bitsch} \& {Izidoro}(2023)}]{bitsch:2023}
{Bitsch}, B., \& {Izidoro}, A. 2023, \aap, 674, A178

\bibitem[{{Bonomo} {et~al.}(2023){Bonomo}, {Dumusque}, {Massa}, {Mortier},
  {Bongiolatti}, {Malavolta}, {Sozzetti}, {Buchhave}, {Damasso}, {Haywood},
  {Morbidelli}, {Latham}, {Molinari}, {Pepe}, {Poretti}, {Udry}, {Affer},
  {Boschin}, {Charbonneau}, {Cosentino}, {Cretignier}, {Ghedina}, {Lega},
  {L{\'o}pez-Morales}, {Margini}, {Mart{\'\i}nez Fiorenzano}, {Mayor},
  {Micela}, {Pedani}, {Pinamonti}, {Rice}, {Sasselov}, {Tronsgaard}, \&
  {Vanderburg}}]{bonomo:2023}
{Bonomo}, A.~S., {Dumusque}, X., {Massa}, A., {et~al.} 2023, \aap, 677, A33

\bibitem[{{Borucki} {et~al.}(2012){Borucki}, {Koch}, {Batalha}, {Bryson},
  {Rowe}, {Fressin}, {Torres}, {Caldwell}, {Christensen-Dalsgaard}, {Cochran},
  {DeVore}, {Gautier}, {Geary}, {Gilliland}, {Gould}, {Howell}, {Jenkins},
  {Latham}, {Lissauer}, {Marcy}, {Sasselov}, {Boss}, {Charbonneau}, {Ciardi},
  {Kaltenegger}, {Doyle}, {Dupree}, {Ford}, {Fortney}, {Holman}, {Steffen},
  {Mullally}, {Still}, {Tarter}, {Ballard}, {Buchhave}, {Carter},
  {Christiansen}, {Demory}, {D{\'e}sert}, {Dressing}, {Endl}, {Fabrycky},
  {Fischer}, {Haas}, {Henze}, {Horch}, {Howard}, {Isaacson}, {Kjeldsen},
  {Johnson}, {Klaus}, {Kolodziejczak}, {Barclay}, {Li}, {Meibom}, {Prsa},
  {Quinn}, {Quintana}, {Robertson}, {Sherry}, {Shporer}, {Tenenbaum},
  {Thompson}, {Twicken}, {Van Cleve}, {Welsh}, {Basu}, {Chaplin}, {Miglio},
  {Kawaler}, {Arentoft}, {Stello}, {Metcalfe}, {Verner}, {Karoff}, {Lundkvist},
  {Lund}, {Handberg}, {Elsworth}, {Hekker}, {Huber}, {Bedding}, \&
  {Rapin}}]{borucki:2012}
{Borucki}, W.~J., {Koch}, D.~G., {Batalha}, N., {et~al.} 2012, \apj, 745, 120

\bibitem[{{Bourrier} {et~al.}(2021){Bourrier}, {Lovis}, {Cretignier}, {Allart},
  {Dumusque}, {Delisle}, {Deline}, {Sousa}, {Adibekyan}, {Alibert}, {Barros},
  {Borsa}, {Cristiani}, {Demangeon}, {Ehrenreich}, {Figueira}, {Gonz{\'a}lez
  Hern{\'a}ndez}, {Lendl}, {Lillo-Box}, {Lo Curto}, {Di Marcantonio},
  {Martins}, {M{\'e}gevand}, {Mehner}, {Micela}, {Molaro}, {Oshagh}, {Palle},
  {Pepe}, {Poretti}, {Rebolo}, {Santos}, {Scandariato}, {Seidel}, {Sozzetti},
  {Su{\'a}rez Mascare{\~n}o}, \& {Zapatero Osorio}}]{bourrier:2021}
{Bourrier}, V., {Lovis}, C., {Cretignier}, M., {et~al.} 2021, \aap, 654, A152

\bibitem[{{Bowler}(2016)}]{bowler:2016}
{Bowler}, B.~P. 2016, \pasp, 128, 102001

\bibitem[{{Brande} {et~al.}(2024){Brande}, {Crossfield}, {Kreidberg}, {Morley},
  {Barman}, {Benneke}, {Christiansen}, {Dragomir}, {Fortney}, {Greene},
  {Hardegree-Ullman}, {Howard}, {Knutson}, {Lothringer}, \&
  {Mikal-Evans}}]{brande:2024}
{Brande}, J., {Crossfield}, I. J.~M., {Kreidberg}, L., {et~al.} 2024, \apjl,
  961, L23

\bibitem[{{Brown} {et~al.}(2013){Brown}, {Baliber}, {Bianco}, {Bowman},
  {Burleson}, {Conway}, {Crellin}, {Depagne}, {De Vera}, {Dilday}, {Dragomir},
  {Dubberley}, {Eastman}, {Elphick}, {Falarski}, {Foale}, {Ford}, {Fulton},
  {Garza}, {Gomez}, {Graham}, {Greene}, {Haldeman}, {Hawkins}, {Haworth},
  {Haynes}, {Hidas}, {Hjelstrom}, {Howell}, {Hygelund}, {Lister}, {Lobdill},
  {Martinez}, {Mullins}, {Norbury}, {Parrent}, {Paulson}, {Petry}, {Pickles},
  {Posner}, {Rosing}, {Ross}, {Sand}, {Saunders}, {Shobbrook}, {Shporer},
  {Street}, {Thomas}, {Tsapras}, {Tufts}, {Valenti}, {Vander Horst}, {Walker},
  {White}, \& {Willis}}]{brown:2013}
{Brown}, T.~M., {Baliber}, N., {Bianco}, F.~B., {et~al.} 2013, Publications of
  the Astronomical Society of the Pacific, 125, 1031

\bibitem[{{Burt} {et~al.}(2018){Burt}, {Holden}, {Wolfgang}, \&
  {Bouma}}]{burt:2018}
{Burt}, J., {Holden}, B., {Wolfgang}, A., \& {Bouma}, L.~G. 2018, \aj, 156, 255

\bibitem[{{Cabona} {et~al.}(2021){Cabona}, {Viana}, {Landoni}, \&
  {Faria}}]{cabona:2021}
{Cabona}, L., {Viana}, P.~T.~P., {Landoni}, M., \& {Faria}, J.~P. 2021, \mnras,
  503, 5504

\bibitem[{{Cale} {et~al.}(2021){Cale}, {Reefe}, {Plavchan}, {Tanner}, {Gaidos},
  {Gagn{\'e}}, {Gao}, {Kane}, {B{\'e}jar}, {Lodieu}, {Anglada-Escud{\'e}},
  {Ribas}, {Pall{\'e}}, {Quirrenbach}, {Amado}, {Reiners}, {Caballero}, {Rosa
  Zapatero Osorio}, {Dreizler}, {Howard}, {Fulton}, {Xuesong Wang}, {Collins},
  {El Mufti}, {Wittrock}, {Gilbert}, {Barclay}, {Klein}, {Martioli},
  {Wittenmyer}, {Wright}, {Addison}, {Hirano}, {Tamura}, {Kotani}, {Narita},
  {Vermilion}, {Lee}, {Geneser}, {Teske}, {Quinn}, {Latham}, {Esquerdo},
  {Calkins}, {Berlind}, {Zohrabi}, {Stibbards}, {Kotnana}, {Jenkins},
  {Twicken}, {Henze}, {Kidwell}, {Burke}, {Villase{\~n}or}, \&
  {Boyd}}]{cale:2021}
{Cale}, B.~L., {Reefe}, M., {Plavchan}, P., {et~al.} 2021, \aj, 162, 295

\bibitem[{{Chontos} {et~al.}(2022){Chontos}, {Murphy}, {MacDougall},
  {Fetherolf}, {Van Zandt}, {Rubenzahl}, {Beard}, {Huber}, {Batalha},
  {Crossfield}, {Dressing}, {Fulton}, {Howard}, {Isaacson}, {Kane}, {Petigura},
  {Robertson}, {Roy}, {Weiss}, {Behmard}, {Dai}, {Dalba}, {Giacalone}, {Hill},
  {Lubin}, {Mayo}, {Mo{\v{c}}nik}, {Polanski}, {Rosenthal}, {Scarsdale},
  {Turtelboom}, {Ricker}, {Vanderspek}, {Latham}, {Seager}, {Winn}, {Jenkins},
  {Quinn}, {Guerrero}, {Collins}, {Ciardi}, {Shporer}, {Goeke}, {Levine},
  {Ting}, {Bieryla}, {Collins}, {Kielkopf}, {Barkaoui}, {Benni},
  {Esparza-Borges}, {Conti}, {Hooton}, {Kagetani}, {Laloum}, {Marino},
  {Massey}, {Murgas}, {Papini}, {Schwarz}, {Srdoc}, {Stockdale}, {Wang},
  {Wittrock}, \& {Zou}}]{chontos:2022}
{Chontos}, A., {Murphy}, J. M.~A., {MacDougall}, M.~G., {et~al.} 2022, \aj,
  163, 297

\bibitem[{{Cloutier} {et~al.}(2019){Cloutier}, {Astudillo-Defru}, {Doyon},
  {Bonfils}, {Almenara}, {Bouchy}, {Delfosse}, {Forveille}, {Lovis}, {Mayor},
  {Menou}, {Murgas}, {Pepe}, {Santos}, {Udry}, \&
  {W{\"u}nsche}}]{cloutier:2019}
{Cloutier}, R., {Astudillo-Defru}, N., {Doyon}, R., {et~al.} 2019, \aap, 621,
  A49

\bibitem[{{Collins}(2019)}]{collins:2019}
{Collins}, K. 2019, in American Astronomical Society Meeting Abstracts, Vol.
  233, American Astronomical Society Meeting Abstracts \#233, 140.05

\bibitem[{{Collins} {et~al.}(2017){Collins}, {Kielkopf}, {Stassun}, \&
  {Hessman}}]{collins:2017}
{Collins}, K.~A., {Kielkopf}, J.~F., {Stassun}, K.~G., \& {Hessman}, F.~V.
  2017, \aj, 153, 77

\bibitem[{{Crass} {et~al.}(2021){Crass}, {Gaudi}, {Leifer}, {Beichman},
  {Bender}, {Blackwood}, {Burt}, {Callas}, {Cegla}, {Diddams}, {Dumusque},
  {Eastman}, {Ford}, {Fulton}, {Gibson}, {Halverson}, {Haywood}, {Hearty},
  {Howard}, {Latham}, {L{\"o}hner-B{\"o}ttcher}, {Mamajek}, {Mortier},
  {Newman}, {Plavchan}, {Quirrenbach}, {Reiners}, {Robertson}, {Roy}, {Schwab},
  {Seifahrt}, {Szentgyorgyi}, {Terrien}, {Teske}, {Thompson}, \&
  {Vasisht}}]{crass:2021}
{Crass}, J., {Gaudi}, B.~S., {Leifer}, S., {et~al.} 2021, arXiv e-prints,
  arXiv:2107.14291

\bibitem[{{Cretignier} {et~al.}(2023){Cretignier}, {Dumusque}, {Aigrain}, \&
  {Pepe}}]{cretignier:2023}
{Cretignier}, M., {Dumusque}, X., {Aigrain}, S., \& {Pepe}, F. 2023, arXiv
  e-prints, arXiv:2308.11812

\bibitem[{{Crossfield}(2023)}]{crossfield:2023}
{Crossfield}, I. J.~M. 2023, \apjl, 952, L18

\bibitem[{{Crossfield} {et~al.}(2015){Crossfield}, {Petigura}, {Schlieder},
  {Howard}, {Fulton}, {Aller}, {Ciardi}, {L{\'e}pine}, {Barclay}, {de Pater},
  {de Kleer}, {Quintana}, {Christiansen}, {Schlafly}, {Kaltenegger}, {Crepp},
  {Henning}, {Obermeier}, {Deacon}, {Weiss}, {Isaacson}, {Hansen}, {Liu},
  {Greene}, {Howell}, {Barman}, \& {Mordasini}}]{crossfield:2015a}
{Crossfield}, I.~J.~M., {Petigura}, E., {Schlieder}, J.~E., {et~al.} 2015,
  \apj, 804, 10

\bibitem[{{Crossfield} {et~al.}(2016){Crossfield}, {Ciardi}, {Petigura},
  {Sinukoff}, {Schlieder}, {Howard}, {Beichman}, {Isaacson}, {Dressing},
  {Christiansen}, {Fulton}, {L{\'e}pine}, {Weiss}, {Hirsch}, {Livingston},
  {Baranec}, {Law}, {Riddle}, {Ziegler}, {Howell}, {Horch}, {Everett}, {Teske},
  {Martinez}, {Obermeier}, {Benneke}, {Scott}, {Deacon}, {Aller}, {Hansen},
  {Mancini}, {Ciceri}, {Brahm}, {Jord{\'a}n}, {Knutson}, {Henning}, {Bonnefoy},
  {Liu}, {Crepp}, {Lothringer}, {Hinz}, {Bailey}, {Skemer}, \&
  {Defrere}}]{crossfield:2016}
{Crossfield}, I.~J.~M., {Ciardi}, D.~R., {Petigura}, E.~A., {et~al.} 2016,
  \apjs, 226, 7

\bibitem[{{Dekany} {et~al.}(2013){Dekany}, {Roberts}, {Burruss}, {Bouchez},
  {Truong}, {Baranec}, {Guiwits}, {Hale}, {Angione}, {Trinh}, {Zolkower},
  {Shelton}, {Palmer}, {Henning}, {Croner}, {Troy}, {McKenna}, {Tesch},
  {Hildebrandt}, \& {Milburn}}]{dekany:2013}
{Dekany}, R., {Roberts}, J., {Burruss}, R., {et~al.} 2013, \apj, 776, 130

\bibitem[{{Deming} \& {Seager}(2017)}]{deming:2017}
{Deming}, L.~D., \& {Seager}, S. 2017, Journal of Geophysical Research
  (Planets), 122, 53

\bibitem[{{Figueira} {et~al.}(2009){Figueira}, {Pont}, {Mordasini}, {Alibert},
  {Georgy}, \& {Benz}}]{figueira:2009}
{Figueira}, P., {Pont}, F., {Mordasini}, C., {et~al.} 2009, \aap, 493, 671

\bibitem[{{Foreman-Mackey} {et~al.}(2013){Foreman-Mackey}, {Hogg}, {Lang}, \&
  {Goodman}}]{foreman-mackey:2012}
{Foreman-Mackey}, D., {Hogg}, D.~W., {Lang}, D., \& {Goodman}, J. 2013, \pasp,
  125, 306

\bibitem[{{Fukui} {et~al.}(2011){Fukui}, {Narita}, {Tristram}, {Sumi}, {Abe},
  {Itow}, {Sullivan}, {Bond}, {Hirano}, {Tamura}, {Bennett}, {Furusawa},
  {Hayashi}, {Hearnshaw}, {Hosaka}, {Kamiya}, {Kobara}, {Korpela}, {Kilmartin},
  {Lin}, {Ling}, {Makita}, {Masuda}, {Matsubara}, {Miyake}, {Muraki}, {Nagaya},
  {Nishimoto}, {Ohnishi}, {Omori}, {Perrott}, {Rattenbury}, {Saito}, {Skuljan},
  {Suzuki}, {Sweatman}, \& {Wada}}]{fukui:2011}
{Fukui}, A., {Narita}, N., {Tristram}, P.~J., {et~al.} 2011, \pasj, 63, 287

\bibitem[{{Fulton} \& {Petigura}(2018)}]{fulton:2018}
{Fulton}, B.~J., \& {Petigura}, E.~A. 2018, \aj, 156, 264

\bibitem[{{Fulton} {et~al.}(2017){Fulton}, {Petigura}, {Howard}, {Isaacson},
  {Marcy}, {Cargile}, {Hebb}, {Weiss}, {Johnson}, {Morton}, {Sinukoff},
  {Crossfield}, \& {Hirsch}}]{fulton:2017}
{Fulton}, B.~J., {Petigura}, E.~A., {Howard}, A.~W., {et~al.} 2017, \aj, 154,
  109

\bibitem[{{Furlan} {et~al.}(2017){Furlan}, {Ciardi}, {Everett}, {Saylors},
  {Teske}, {Horch}, {Howell}, {van Belle}, {Hirsch}, {Gautier}, {Adams},
  {Barrado}, {Cartier}, {Dressing}, {Dupree}, {Gilliland}, {Lillo-Box},
  {Lucas}, \& {Wang}}]{furlan:2017}
{Furlan}, E., {Ciardi}, D.~R., {Everett}, M.~E., {et~al.} 2017, \aj, 153, 71

\bibitem[{{Galland} {et~al.}(2005){Galland}, {Lagrange}, {Udry}, {Chelli},
  {Pepe}, {Queloz}, {Beuzit}, \& {Mayor}}]{galland:2005}
{Galland}, F., {Lagrange}, A.-M., {Udry}, S., {et~al.} 2005, \aap, 443, 337

\bibitem[{{Gavel} {et~al.}(2014){Gavel}, {Kupke}, {Dillon}, {Norton},
  {Ratliff}, {Cabak}, {Phillips}, {Rockosi}, {McGurk}, {Srinath}, {Peck},
  {Deich}, {Lanclos}, {Gates}, {Saylor}, {Ward}, \& {Pfister}}]{gavel:2014}
{Gavel}, D., {Kupke}, R., {Dillon}, D., {et~al.} 2014, in Society of
  Photo-Optical Instrumentation Engineers (SPIE) Conference Series, Vol. 9148,
  Adaptive Optics Systems IV, ed. E.~{Marchetti}, L.~M. {Close}, \& J.-P.
  {Vran}, 914805

\bibitem[{{Giacalone} {et~al.}(2021){Giacalone}, {Dressing}, {Jensen},
  {Collins}, {Ricker}, {Vanderspek}, {Seager}, {Winn}, {Jenkins}, {Barclay},
  {Barkaoui}, {Cadieux}, {Charbonneau}, {Collins}, {Conti}, {Doyon}, {Evans},
  {Ghachoui}, {Gillon}, {Guerrero}, {Hart}, {Jehin}, {Kielkopf}, {McLean},
  {Murgas}, {Palle}, {Parviainen}, {Pozuelos}, {Relles}, {Shporer}, {Socia},
  {Stockdale}, {Tan}, {Torres}, {Twicken}, {Waalkes}, \&
  {Waite}}]{giacalone:2021}
{Giacalone}, S., {Dressing}, C.~D., {Jensen}, E. L.~N., {et~al.} 2021, \aj,
  161, 24

\bibitem[{{Gibson} {et~al.}(2016){Gibson}, {Howard}, {Marcy}, {Edelstein},
  {Wishnow}, \& {Poppett}}]{gibson:2016}
{Gibson}, S.~R., {Howard}, A.~W., {Marcy}, G.~W., {et~al.} 2016, in Society of
  Photo-Optical Instrumentation Engineers (SPIE) Conference Series, Vol. 9908,
  Ground-based and Airborne Instrumentation for Astronomy VI, ed. C.~J.
  {Evans}, L.~{Simard}, \& H.~{Takami}, 990870

\bibitem[{{Gilbert} \& {Petigura}(2024)}]{gilbert:2024}
{Gilbert}, G., \& {Petigura}, E. 2024, in AAS/Division for Extreme Solar
  Systems Abstracts, Vol.~56, AAS/Division for Extreme Solar Systems Abstracts,
  301.05

\bibitem[{{Hayward} {et~al.}(2001){Hayward}, {Brandl}, {Pirger}, {Blacken},
  {Gull}, {Schoenwald}, \& {Houck}}]{hayward:2001}
{Hayward}, T.~L., {Brandl}, B., {Pirger}, B., {et~al.} 2001, \pasp, 113, 105

\bibitem[{{Hejazi} {et~al.}(2023){Hejazi}, {Crossfield}, {Nordlander},
  {Mansfield}, {Souto}, {Marfil}, {Coria}, {Brande}, {Polanski}, {Hand}, \&
  {Wienke}}]{hejazi:2023}
{Hejazi}, N., {Crossfield}, I. J.~M., {Nordlander}, T., {et~al.} 2023, \apj,
  949, 79

\bibitem[{{Hodapp} {et~al.}(2003){Hodapp}, {Irwin}, {Yamada}, {Chung},
  {Fletcher}, {Jensen}, {Mays}, {Nolan}, {Simons}, \& {Aspin}}]{hodapp:2003}
{Hodapp}, K.-W., {Irwin}, E.~M., {Yamada}, H., {et~al.} 2003, in \procspie,
  Vol. 4841, Instrument Design and Performance for Optical/Infrared
  Ground-based Telescopes, ed. M.~{Iye} \& A.~F.~M. {Moorwood}, 869--880

\bibitem[{{Howard} {et~al.}(2010){Howard}, {Marcy}, {Johnson}, {Fischer},
  {Wright}, {Isaacson}, {Valenti}, {Anderson}, {Lin}, \& {Ida}}]{howard:2010}
{Howard}, A.~W., {Marcy}, G.~W., {Johnson}, J.~A., {et~al.} 2010, Science, 330,
  653

\bibitem[{{Howell} {et~al.}(2011){Howell}, {Everett}, {Sherry}, {Horch}, \&
  {Ciardi}}]{howell:2011}
{Howell}, S.~B., {Everett}, M.~E., {Sherry}, W., {Horch}, E., \& {Ciardi},
  D.~R. 2011, \aj, 142, 19

\bibitem[{{Jenkins} {et~al.}(2016){Jenkins}, {Twicken}, {McCauliff},
  {Campbell}, {Sanderfer}, {Lung}, {Mansouri-Samani}, {Girouard}, {Tenenbaum},
  \& {Klaus}}]{jenkins:2016}
{Jenkins}, J.~M., {Twicken}, J.~D., {McCauliff}, S., {et~al.} 2016, in Society
  of Photo-Optical Instrumentation Engineers (SPIE) Conference Series, Vol.
  9913, Software and Cyberinfrastructure for Astronomy IV, 99133E

\bibitem[{{Jensen}(2013)}]{Jensen:2013}
{Jensen}, E. 2013, {Tapir: A web interface for transit/eclipse observability},
  Astrophysics Source Code Library, ascl:1306.007

\bibitem[{{Johnson} {et~al.}(2017){Johnson}, {Petigura}, {Fulton}, {Marcy},
  {Howard}, {Isaacson}, {Hebb}, {Cargile}, {Morton}, {Weiss}, {Winn}, {Rogers},
  {Sinukoff}, \& {Hirsch}}]{johnson:2017}
{Johnson}, J.~A., {Petigura}, E.~A., {Fulton}, B.~J., {et~al.} 2017, \aj, 154,
  108

\bibitem[{{JTEC Team} {et~al.}(2023){JTEC Team}, {Ahrer}, {Alderson},
  {Batalha}, {Batalha}, {Bean}, {Beatty}, {Bell}, {Benneke}, {Berta-Thompson},
  {Carter}, {Crossfield}, {Espinoza}, {Feinstein}, {Fortney}, {Gibson},
  {Goyal}, {Kempton}, {Kirk}, {Kreidberg}, {L{\'o}pez-Morales}, {Line},
  {Lothringer}, {Moran}, {Mukherjee}, {Ohno}, {Parmentier}, {Piaulet},
  {Rustamkulov}, {Schlawin}, {Sing}, {Stevenson}, {Wakeford}, {Allen},
  {Birkmann}, {Brande}, {Crouzet}, {Cubillos}, {Damiano}, {D{\'e}sert}, {Gao},
  {Harrington}, {Hu}, {Kendrew}, {Knutson}, {Lagage}, {Leconte}, {Lendl},
  {MacDonald}, {May}, {Miguel}, {Molaverdikhani}, {Moses}, {Murray}, {Nehring},
  {Nikolov}, {Petit dit de la Roche}, {Radica}, {Roy}, {Stassun}, {Taylor},
  {Waalkes}, {Wachiraphan}, {Welbanks}, {Wheatley}, {Aggarwal}, {Alam},
  {Banerjee}, {Barstow}, {Blecic}, {Casewell}, {Changeat}, {Chubb},
  {Col{\'o}n}, {Coulombe}, {Daylan}, {de Val-Borro}, {Decin}, {Dos Santos},
  {Flagg}, {France}, {Fu}, {Garc{\'\i}a Mu{\~n}oz}, {Gizis}, {Glidden},
  {Grant}, {Heng}, {Henning}, {Hong}, {Inglis}, {Iro}, {Kataria}, {Komacek},
  {Krick}, {Lee}, {Lewis}, {Lillo-Box}, {Lustig-Yaeger}, {Mancini}, {Mandell},
  {Mansfield}, {Marley}, {Mikal-Evans}, {Morello}, {Nixon}, {Ortiz Ceballos},
  {Piette}, {Powell}, {Rackham}, {Ramos-Rosado}, {Rauscher}, {Redfield},
  {Rogers}, {Roman}, {Roudier}, {Scarsdale}, {Shkolnik}, {Southworth}, {Spake},
  {Steinrueck}, {Tan}, {Teske}, {Tremblin}, {Tsai}, {Tucker}, {Turner},
  {Valenti}, {Venot}, {Waldmann}, {Wallack}, {Zhang}, \& {Zieba}}]{ers:2023}
{JTEC Team}, {Ahrer}, E.-M., {Alderson}, L., {et~al.} 2023, \nat, 614, 649

\bibitem[{{Kanodia} {et~al.}(2023){Kanodia}, {He}, {Ford}, {Ghosh}, \&
  {Wolfgang}}]{kanodia:2023}
{Kanodia}, S., {He}, M.~Y., {Ford}, E.~B., {Ghosh}, S.~K., \& {Wolfgang}, A.
  2023, \apj, 956, 76

\bibitem[{{Kanodia} {et~al.}(2019){Kanodia}, {Wolfgang}, {Stefansson}, {Ning},
  \& {Mahadevan}}]{kanodia:2019}
{Kanodia}, S., {Wolfgang}, A., {Stefansson}, G.~K., {Ning}, B., \& {Mahadevan},
  S. 2019, \apj, 882, 38

\bibitem[{{Kempton} {et~al.}(2018){Kempton}, {Bean}, {Louie}, {Deming}, {Koll},
  {Mansfield}, {Christiansen}, {L{\'o}pez-Morales}, {Swain}, {Zellem},
  {Ballard}, {Barclay}, {Barstow}, {Batalha}, {Beatty}, {Berta-Thompson},
  {Birkby}, {Buchhave}, {Charbonneau}, {Cowan}, {Crossfield}, {de Val-Borro},
  {Doyon}, {Dragomir}, {Gaidos}, {Heng}, {Hu}, {Kane}, {Kreidberg}, {Mallonn},
  {Morley}, {Narita}, {Nascimbeni}, {Pall{\'e}}, {Quintana}, {Rauscher},
  {Seager}, {Shkolnik}, {Sing}, {Sozzetti}, {Stassun}, {Valenti}, \& {von
  Essen}}]{kempton:2018}
{Kempton}, E. M.~R., {Bean}, J.~L., {Louie}, D.~R., {et~al.} 2018, \pasp, 130,
  114401

\bibitem[{{Kolecki} \& {Wang}(2022)}]{kolecki:2022}
{Kolecki}, J.~R., \& {Wang}, J. 2022, \aj, 164, 87

\bibitem[{{Konopacky} {et~al.}(2013){Konopacky}, {Barman}, {Macintosh}, \&
  {Marois}}]{konopacky:2013}
{Konopacky}, Q.~M., {Barman}, T.~S., {Macintosh}, B.~A., \& {Marois}, C. 2013,
  Science, 339, 1398

\bibitem[{{Kreidberg}(2015)}]{kreidberg:2015b}
{Kreidberg}, L. 2015, \pasp, 127, 1161

\bibitem[{{Kupke} {et~al.}(2012){Kupke}, {Gavel}, {Roskosi}, {Cabak}, {Cowley},
  {Dillon}, {Gates}, {McGurk}, {Norton}, {Peck}, {Ratliff}, \&
  {Reinig}}]{kupke:2012}
{Kupke}, R., {Gavel}, D., {Roskosi}, C., {et~al.} 2012, in Society of
  Photo-Optical Instrumentation Engineers (SPIE) Conference Series, Vol. 8447,
  Adaptive Optics Systems III, ed. B.~L. {Ellerbroek}, E.~{Marchetti}, \& J.-P.
  {V{\'e}ran}, 84473G

\bibitem[{{Libby-Roberts} {et~al.}(2020){Libby-Roberts}, {Berta-Thompson},
  {D{\'e}sert}, {Masuda}, {Morley}, {Lopez}, {Deck}, {Fabrycky}, {Fortney},
  {Line}, {Sanchis-Ojeda}, \& {Winn}}]{libby:2020}
{Libby-Roberts}, J.~E., {Berta-Thompson}, Z.~K., {D{\'e}sert}, J.-M., {et~al.}
  2020, \aj, 159, 57

\bibitem[{{Lightkurve Collaboration} {et~al.}(2018){Lightkurve Collaboration},
  {Cardoso}, {Hedges}, {Gully-Santiago}, {Saunders}, {Cody}, {Barclay}, {Hall},
  {Sagear}, {Turtelboom}, {Zhang}, {Tzanidakis}, {Mighell}, {Coughlin}, {Bell},
  {Berta-Thompson}, {Williams}, {Dotson}, \& {Barentsen}}]{lightkurve:2018}
{Lightkurve Collaboration}, {Cardoso}, J. V. d.~M., {Hedges}, C., {et~al.}
  2018, {Lightkurve: Kepler and TESS time series analysis in Python},
  Astrophysics Source Code Library, record ascl:1812.013, ascl:1812.013

\bibitem[{{Lissauer} {et~al.}(2011){Lissauer}, {Ragozzine}, {Fabrycky},
  {Steffen}, {Ford}, {Jenkins}, {Shporer}, {Holman}, {Rowe}, {Quintana},
  {Batalha}, {Borucki}, {Bryson}, {Caldwell}, {Carter}, {Ciardi}, {Dunham},
  {Fortney}, {Gautier}, {Howell}, {Koch}, {Latham}, {Marcy}, {Morehead}, \&
  {Sasselov}}]{lissauer:2011a}
{Lissauer}, J.~J., {Ragozzine}, D., {Fabrycky}, D.~C., {et~al.} 2011, \apjs,
  197, 8

\bibitem[{{Lopez} \& {Fortney}(2014)}]{lopez:2014}
{Lopez}, E.~D., \& {Fortney}, J.~J. 2014, \apj, 792, 1

\bibitem[{{Lothringer} {et~al.}(2021){Lothringer}, {Rustamkulov}, {Sing},
  {Gibson}, {Wilson}, \& {Schlaufman}}]{lothringer:2021}
{Lothringer}, J.~D., {Rustamkulov}, Z., {Sing}, D.~K., {et~al.} 2021, \apj,
  914, 12

\bibitem[{{Luhn} {et~al.}(2023){Luhn}, {Ford}, {Guo}, {Gilbertson}, {Newman},
  {Plavchan}, {Burt}, {Teske}, \& {Gupta}}]{luhn:2023}
{Luhn}, J.~K., {Ford}, E.~B., {Guo}, Z., {et~al.} 2023, \aj, 165, 98

\bibitem[{{Luque} \& {Pall{\'e}}(2022)}]{luque:2022}
{Luque}, R., \& {Pall{\'e}}, E. 2022, Science, 377, 1211

\bibitem[{{Madhusudhan} {et~al.}(2023){Madhusudhan}, {Sarkar}, {Constantinou},
  {Holmberg}, {Piette}, \& {Moses}}]{madhusudhan:2023b}
{Madhusudhan}, N., {Sarkar}, S., {Constantinou}, S., {et~al.} 2023, arXiv
  e-prints, arXiv:2309.05566

\bibitem[{{Mamajek} \& {Stapelfeldt}(2024)}]{mamajek:2024}
{Mamajek}, E., \& {Stapelfeldt}, K. 2024, arXiv e-prints, arXiv:2402.12414

\bibitem[{{McCully} {et~al.}(2018){McCully}, {Volgenau}, {Harbeck}, {Lister},
  {Saunders}, {Turner}, {Siiverd}, \& {Bowman}}]{mccully:2018}
{McCully}, C., {Volgenau}, N.~H., {Harbeck}, D.-R., {et~al.} 2018, in Society
  of Photo-Optical Instrumentation Engineers (SPIE) Conference Series, Vol.
  10707, \procspie, 107070K

\bibitem[{{McGurk} {et~al.}(2014){McGurk}, {Rockosi}, {Gavel}, {Kupke}, {Peck},
  {Pfister}, {Ward}, {Deich}, {Gates}, {Gates}, {Alcott}, {Cowley}, {Dillon},
  {Lanclos}, {Sandford}, {Saylor}, {Srinath}, {Weiss}, \&
  {Norton}}]{mcgurk:2014}
{McGurk}, R., {Rockosi}, C., {Gavel}, D., {et~al.} 2014, in Society of
  Photo-Optical Instrumentation Engineers (SPIE) Conference Series, Vol. 9148,
  Adaptive Optics Systems IV, ed. E.~{Marchetti}, L.~M. {Close}, \& J.-P.
  {Vran}, 91483A

\bibitem[{{Millholland} {et~al.}(2022){Millholland}, {He}, \&
  {Zink}}]{millholland:2022}
{Millholland}, S.~C., {He}, M.~Y., \& {Zink}, J.~K. 2022, \aj, 164, 72

\bibitem[{{Naponiello} {et~al.}(2023){Naponiello}, {Mancini}, {Sozzetti},
  {Bonomo}, {Morbidelli}, {Dou}, {Zeng}, {Leinhardt}, {Biazzo}, {Cubillos},
  {Pinamonti}, {Locci}, {Maggio}, {Damasso}, {Lanza}, {Lissauer}, {Bignamini},
  {Boschin}, {Bouma}, {Carter}, {Ciardi}, {Collins}, {Cosentino}, {Crossfield},
  {Desidera}, {Dumusque}, {Fiorenzano}, {Fukui}, {Giacobbe}, {Gnilka},
  {Ghedina}, {Gonzales}, {Guilluy}, {Harutyunyan}, {Howell}, {Jenkins}, {Lund},
  {Jensen}, {Kielkopf}, {Lester}, {Malavolta}, {Mann}, {Matson}, {Matthews},
  {Nardiello}, {Narita}, {Pace}, {Pagano}, {Palle}, {Pedani}, {Seager},
  {Schlieder}, {Schwarz}, {Shporer}, {Twicken}, {Winn}, {Ziegler}, \&
  {Zingales}}]{naponiello:2023}
{Naponiello}, L., {Mancini}, L., {Sozzetti}, A., {et~al.} 2023, arXiv e-prints,
  arXiv:2309.01464

\bibitem[{{Narita} {et~al.}(2015){Narita}, {Fukui}, {Kusakabe}, {Onitsuka},
  {Ryu}, {Yanagisawa}, {Izumiura}, {Tamura}, \& {Yamamuro}}]{narita:2015a}
{Narita}, N., {Fukui}, A., {Kusakabe}, N., {et~al.} 2015, Journal of
  Astronomical Telescopes, Instruments, and Systems, 1, 045001

\bibitem[{Narita {et~al.}(2019)Narita, Fukui, Kusakabe, Watanabe, Palle,
  Parviainen, Monta{\~{n}}{\'{e}}s-Rodr{\'{i}}guez, Murgas, Monelli, Aguiar, \&
  {Perez Prieto}}]{narita:2019}
Narita, N., Fukui, A., Kusakabe, N., {et~al.} 2019, J. Astron. Telesc.
  Instruments, Syst., 5, 015001

\bibitem[{{Narita} {et~al.}(2020){Narita}, {Fukui}, {Yamamuro}, {Harbeck},
  {Bowman}, {Elphick}, {Nation}, {Armstrong}, {Han}, {Abe}, {Ikoma}, {Isogai},
  {Kawauchi}, {Kurita}, {Kusakabe}, {de Leon}, {Livingston}, {Mori},
  {Nishiumi}, {Tamura}, {Watanabe}, {Volgenau}, {Heinrich-Josties}, {Foale},
  {Daily}, {McCully}, {Kirby}, {Smith}, {Haworth}, {Conway},
  {Storrie-Lombardi}, {Rosing}, {Chatelain}, {Bachelet}, {Johnson}, \&
  {Rabus}}]{narita:2020}
{Narita}, N., {Fukui}, A., {Yamamuro}, T., {et~al.} 2020, in Society of
  Photo-Optical Instrumentation Engineers (SPIE) Conference Series, Vol. 11447,
  Society of Photo-Optical Instrumentation Engineers (SPIE) Conference Series,
  114475K

\bibitem[{{Ning} {et~al.}(2018){Ning}, {Wolfgang}, \& {Ghosh}}]{ning:2018}
{Ning}, B., {Wolfgang}, A., \& {Ghosh}, S. 2018, \apj, 869, 5

\bibitem[{{Otegi} {et~al.}(2020){Otegi}, {Bouchy}, \& {Helled}}]{otegi:2020}
{Otegi}, J.~F., {Bouchy}, F., \& {Helled}, R. 2020, \aap, 634, A43

\bibitem[{{Ould-Elhkim} {et~al.}(2023){Ould-Elhkim}, {Moutou}, {Donati},
  {Artigau}, {Fouqu{\'e}}, {Cook}, {Carmona}, {Cristofari}, {Martioli},
  {Debras}, {Dumusque}, {Martins}, {H{\'e}brard}, {Cadieux}, {Delfosse},
  {Doyon}, {Klein}, {Gomes da Silva}, {Forveille}, {Hood}, \&
  {Charpentier}}]{ould-elhkim:2023}
{Ould-Elhkim}, M., {Moutou}, C., {Donati}, J.~F., {et~al.} 2023, \aap, 675,
  A187

\bibitem[{{Owen}(2019)}]{owen:2019}
{Owen}, J.~E. 2019, Annual Review of Earth and Planetary Sciences, 47, 67

\bibitem[{{Parc} {et~al.}(2024){Parc}, {Bouchy}, {Venturini}, {Dorn}, \&
  {Helled}}]{parc:2024}
{Parc}, L., {Bouchy}, F., {Venturini}, J., {Dorn}, C., \& {Helled}, R. 2024,
  \aap, 688, A59

\bibitem[{{Parviainen} \& {Aigrain}(2015)}]{parviainen:2015}
{Parviainen}, H., \& {Aigrain}, S. 2015, \mnras, 453, 3821

\bibitem[{{Parviainen} {et~al.}(2024){Parviainen}, {Luque}, \&
  {Palle}}]{parviainen:2024}
{Parviainen}, H., {Luque}, R., \& {Palle}, E. 2024, \mnras, 527, 5693

\bibitem[{{Parviainen} {et~al.}(2019){Parviainen}, {Tingley}, {Deeg}, {Palle},
  {Alonso}, {Montanes Rodriguez}, {Murgas}, {Narita}, {Fukui}, {Watanabe},
  {Kusakabe}, {Tamura}, {Nishiumi}, {Prieto-Arranz}, {Klagyivik}, {B{\'e}jar},
  {Crouzet}, {Mori}, {Hidalgo Soto}, {Casasayas Barris}, \&
  {Luque}}]{parviainen:2020}
{Parviainen}, H., {Tingley}, B., {Deeg}, H.~J., {et~al.} 2019, \aap, 630, A89

\bibitem[{{Petigura} {et~al.}(2017){Petigura}, {Howard}, {Marcy}, {Johnson},
  {Isaacson}, {Cargile}, {Hebb}, {Fulton}, {Weiss}, {Morton}, {Winn}, {Rogers},
  {Sinukoff}, {Hirsch}, \& {Crossfield}}]{petigura:2017}
{Petigura}, E.~A., {Howard}, A.~W., {Marcy}, G.~W., {et~al.} 2017, \aj, 154,
  107

\bibitem[{{Polanski} {et~al.}(2022){Polanski}, {Crossfield}, {Howard},
  {Isaacson}, \& {Rice}}]{polanski:2022}
{Polanski}, A.~S., {Crossfield}, I. J.~M., {Howard}, A.~W., {Isaacson}, H., \&
  {Rice}, M. 2022, Research Notes of the American Astronomical Society, 6, 155

\bibitem[{{Polanski} {et~al.}(2024){Polanski}, {Lubin}, {Beard}, {Akana
  Murphy}, {Rubenzahl}, {Hill}, {Crossfield}, {Chontos}, {Robertson},
  {Isaacson}, {Kane}, {Ciardi}, {Batalha}, {Dressing}, {Fulton}, {Howard},
  {Huber}, {Petigura}, {Weiss}, {Angelo}, {Behmard}, {Blunt}, {Brinkman},
  {Dai}, {Dalba}, {Fetherolf}, {Giacalone}, {Hirsch}, {Holcomb}, {Kosiarek},
  {Mayo}, {MacDougall}, {Mo{\v{c}}nik}, {Pidhorodetska}, {Rice}, {Rosenthal},
  {Scarsdale}, {Turtelboom}, {Tyler}, {Van Zandt}, {Yee}, {Coria}, {Dulz},
  {Hartman}, {Householder}, {Lange}, {Langford}, {Louden}, {Siegel}, {Gilbert},
  {Gonzales}, {Schlieder}, {Boyle}, {Christiansen}, {Clark}, {Fernandes},
  {Lund}, {Savel}, {Gill}, {Beichman}, {Matson}, {Matthews}, {Furlan},
  {Howell}, {Scott}, {Everett}, {Livingston}, {Ershova}, {Cheryasov},
  {Safonov}, {Lillo-Box}, {Barrado}, \& {Morales-Calder{\'o}n}}]{polanski:2024}
{Polanski}, A.~S., {Lubin}, J., {Beard}, C., {et~al.} 2024, \apjs, 272, 32

\bibitem[{{Radica} {et~al.}(2022){Radica}, {Artigau}, {Lafreni{\'e}re},
  {Cadieux}, {Cook}, {Doyon}, {Amado}, {Caballero}, {Henning}, {Quirrenbach},
  {Reiners}, \& {Ribas}}]{radica:2022}
{Radica}, M., {Artigau}, {\'E}., {Lafreni{\'e}re}, D., {et~al.} 2022, \mnras,
  517, 5050

\bibitem[{{Rogers} {et~al.}(2023){Rogers}, {Schlichting}, \&
  {Owen}}]{rogers:2023}
{Rogers}, J.~G., {Schlichting}, H.~E., \& {Owen}, J.~E. 2023, \apjl, 947, L19

\bibitem[{{Savel} {et~al.}(2020){Savel}, {Dressing}, {Hirsch}, {Ciardi},
  {Fleming}, {Giacalone}, {Mayo}, \& {Christiansen}}]{savel:2020}
{Savel}, A.~B., {Dressing}, C.~D., {Hirsch}, L.~A., {et~al.} 2020, \aj, 160,
  287

\bibitem[{{Savel} {et~al.}(2022){Savel}, {Hirsch}, {Gill}, {Dressing}, \&
  {Ciardi}}]{savel:2022}
{Savel}, A.~B., {Hirsch}, L.~A., {Gill}, H., {Dressing}, C.~D., \& {Ciardi},
  D.~R. 2022, \pasp, 134, 124501

\bibitem[{{Schlieder} {et~al.}(2021){Schlieder}, {Gonzales}, {Ciardi}, {Patel},
  {Crossfield}, {Crepp}, {Dressing}, {Barclay}, \& {Howard}}]{schlieder:2021}
{Schlieder}, J.~E., {Gonzales}, E.~J., {Ciardi}, D.~R., {et~al.} 2021,
  Frontiers in Astronomy and Space Sciences, 8, 63

\bibitem[{{Schwab} {et~al.}(2016){Schwab}, {Rakich}, {Gong}, {Mahadevan},
  {Halverson}, {Roy}, {Terrien}, {Robertson}, {Hearty}, {Levi}, {Monson},
  {Wright}, {McElwain}, {Bender}, {Blake}, {St{\"u}rmer}, {Gurevich},
  {Chakraborty}, \& {Ramsey}}]{schwab:2016b}
{Schwab}, C., {Rakich}, A., {Gong}, Q., {et~al.} 2016, in Society of
  Photo-Optical Instrumentation Engineers (SPIE) Conference Series, Vol. 9908,
  Ground-based and Airborne Instrumentation for Astronomy VI, ed. C.~J.
  {Evans}, L.~{Simard}, \& H.~{Takami}, 99087H

\bibitem[{{Scott} \& {Howell}(2018)}]{scott:2018}
{Scott}, N.~J., \& {Howell}, S.~B. 2018, in Society of Photo-Optical
  Instrumentation Engineers (SPIE) Conference Series, Vol. 10701, Optical and
  Infrared Interferometry and Imaging VI, 107010G

\bibitem[{{Sing} {et~al.}(2024){Sing}, {Rustamkulov}, {Thorngren}, {Barstow},
  {Tremblin}, {Alves de Oliveira}, {Beck}, {Birkmann}, {Challener}, {Crouzet},
  {Espinoza}, {Ferruit}, {Giardino}, {Gressier}, {Lee}, {Lewis}, {Maiolino},
  {Manjavacas}, {Rauscher}, {Sirianni}, \& {Valenti}}]{sing:2024}
{Sing}, D.~K., {Rustamkulov}, Z., {Thorngren}, D.~P., {et~al.} 2024, arXiv
  e-prints, arXiv:2405.11027

\bibitem[{{Stassun} {et~al.}(2018){Stassun}, {Oelkers}, {Pepper}, {Paegert},
  {De Lee}, {Torres}, {Latham}, {Charpinet}, {Dressing}, \&
  {Huber}}]{stassun:2018b}
{Stassun}, K.~G., {Oelkers}, R.~J., {Pepper}, J., {et~al.} 2018, \aj, 156, 102

\bibitem[{{Teske} {et~al.}(2021){Teske}, {Wang}, {Wolfgang}, {Gan},
  {Plotnykov}, {Armstrong}, {Butler}, {Cale}, {Crane}, {Howard}, {Jensen},
  {Law}, {Shectman}, {Plavchan}, {Valencia}, {Vanderburg}, {Ricker},
  {Vanderspek}, {Latham}, {Seager}, {Winn}, {Jenkins}, {Adibekyan}, {Barrado},
  {Barros}, {Benkhaldoun}, {Brown}, {Bryant}, {Burt}, {Caldwell},
  {Charbonneau}, {Cloutier}, {Collins}, {Collins}, {Colon}, {Conti},
  {Demangeon}, {Eastman}, {Elmufti}, {Feng}, {Flowers}, {Guerrero},
  {Hojjatpanah}, {Irwin}, {Isopi}, {Lillo-Box}, {Mallia}, {Massey}, {Mori},
  {Mullally}, {Narita}, {Nishiumi}, {Osborn}, {Paegert}, {de Leon}, {Quinn},
  {Reefe}, {Schwarz}, {Shporer}, {Soubkiou}, {Sousa}, {Stockdale}, {Str{\o}m},
  {Tan}, {Tang}, {Tenenbaum}, {Wheatley}, {Wittrock}, {Yahalomi}, \&
  {Zohrabi}}]{teske:2021}
{Teske}, J., {Wang}, S.~X., {Wolfgang}, A., {et~al.} 2021, \apjs, 256, 33

\bibitem[{{Tinetti} {et~al.}(2018){Tinetti}, {Drossart}, {Eccleston},
  {Hartogh}, {Heske}, {Leconte}, {Micela}, {Ollivier}, {Pilbratt}, {Puig},
  {Turrini}, {Vandenbussche}, {Wolkenberg}, {Beaulieu}, {Buchave}, {Ferus},
  {Griffin}, {Guedel}, {Justtanont}, {Lagage}, {Machado}, {Malaguti}, {Min},
  {N{\o}rgaard-Nielsen}, {Rataj}, {Ray}, {Ribas}, {Swain}, {Szabo}, {Werner},
  {Barstow}, {Burleigh}, {Cho}, {Coud{\'e} du Foresto}, {Coustenis}, {Decin},
  {Encrenaz}, {Galand}, {Gillon}, {Helled}, {Morales}, {Garc{\'\i}a Mu{\~n}oz},
  {Moneti}, {Pagano}, {Pascale}, {Piccioni}, {Pinfield}, {Sarkar}, {Selsis},
  {Tennyson}, {Triaud}, {Venot}, {Waldmann}, {Waltham}, {Wright}, {Amiaux},
  {Augu{\`e}res}, {Berth{\'e}}, {Bezawada}, {Bishop}, {Bowles}, {Coffey},
  {Colom{\'e}}, {Crook}, {Crouzet}, {Da Peppo}, {Sanz}, {Focardi}, {Frericks},
  {Hunt}, {Kohley}, {Middleton}, {Morgante}, {Ottensamer}, {Pace}, {Pearson},
  {Stamper}, {Symonds}, {Rengel}, {Renotte}, {Ade}, {Affer}, {Alard}, {Allard},
  {Altieri}, {Andr{\'e}}, {Arena}, {Argyriou}, {Aylward}, {Baccani}, {Bakos},
  {Banaszkiewicz}, {Barlow}, {Batista}, {Bellucci}, {Benatti}, {Bernardi},
  {B{\'e}zard}, {Blecka}, {Bolmont}, {Bonfond}, {Bonito}, {Bonomo}, {Brucato},
  {Brun}, {Bryson}, {Bujwan}, {Casewell}, {Charnay}, {Pestellini}, {Chen},
  {Ciaravella}, {Claudi}, {Cl{\'e}dassou}, {Damasso}, {Damiano}, {Danielski},
  {Deroo}, {Di Giorgio}, {Dominik}, {Doublier}, {Doyle}, {Doyon}, {Drummond},
  {Duong}, {Eales}, {Edwards}, {Farina}, {Flaccomio}, {Fletcher}, {Forget},
  {Fossey}, {Fr{\"a}nz}, {Fujii}, {Garc{\'\i}a-Piquer}, {Gear}, {Geoffray},
  {G{\'e}rard}, {Gesa}, {Gomez}, {Graczyk}, {Griffith}, {Grodent}, {Guarcello},
  {Gustin}, {Hamano}, {Hargrave}, {Hello}, {Heng}, {Herrero}, {Hornstrup},
  {Hubert}, {Ida}, {Ikoma}, {Iro}, {Irwin}, {Jarchow}, {Jaubert}, {Jones},
  {Julien}, {Kameda}, {Kerschbaum}, {Kervella}, {Koskinen}, {Krijger}, {Krupp},
  {Lafarga}, {Landini}, {Lellouch}, {Leto}, {Luntzer}, {Rank-L{\"u}ftinger},
  {Maggio}, {Maldonado}, {Maillard}, {Mall}, {Marquette}, {Mathis}, {Maxted},
  {Matsuo}, {Medvedev}, {Miguel}, {Minier}, {Morello}, {Mura}, {Narita},
  {Nascimbeni}, {Nguyen Tong}, {Noce}, {Oliva}, {Palle}, {Palmer}, {Pancrazzi},
  {Papageorgiou}, {Parmentier}, {Perger}, {Petralia}, {Pezzuto},
  {Pierrehumbert}, {Pillitteri}, {Piotto}, {Pisano}, {Prisinzano}, {Radioti},
  {R{\'e}ess}, {Rezac}, {Rocchetto}, {Rosich}, {Sanna}, {Santerne}, {Savini},
  {Scandariato}, {Sicardy}, {Sierra}, {Sindoni}, {Skup}, {Snellen}, {Sobiecki},
  {Soret}, {Sozzetti}, {Stiepen}, {Strugarek}, {Taylor}, {Taylor}, {Terenzi},
  {Tessenyi}, {Tsiaras}, {Tucker}, {Valencia}, {Vasisht}, {Vazan}, {Vilardell},
  {Vinatier}, {Viti}, {Waters}, {Wawer}, {Wawrzaszek}, {Whitworth}, {Yung},
  {Yurchenko}, {Zapatero Osorio}, {Zellem}, {Zingales}, \&
  {Zwart}}]{tinetti:2018}
{Tinetti}, G., {Drossart}, P., {Eccleston}, P., {et~al.} 2018, Experimental
  Astronomy, 46, 135

\bibitem[{{Tsai} {et~al.}(2023){Tsai}, {Lee}, {Powell}, {Gao}, {Zhang},
  {Moses}, {H{\'e}brard}, {Venot}, {Parmentier}, {Jordan}, {Hu}, {Alam},
  {Alderson}, {Batalha}, {Bean}, {Benneke}, {Bierson}, {Brady}, {Carone},
  {Carter}, {Chubb}, {Inglis}, {Leconte}, {Line}, {L{\'o}pez-Morales},
  {Miguel}, {Molaverdikhani}, {Rustamkulov}, {Sing}, {Stevenson}, {Wakeford},
  {Yang}, {Aggarwal}, {Baeyens}, {Barat}, {de Val-Borro}, {Daylan}, {Fortney},
  {France}, {Goyal}, {Grant}, {Kirk}, {Kreidberg}, {Louca}, {Moran},
  {Mukherjee}, {Nasedkin}, {Ohno}, {Rackham}, {Redfield}, {Taylor}, {Tremblin},
  {Visscher}, {Wallack}, {Welbanks}, {Youngblood}, {Ahrer}, {Batalha}, {Behr},
  {Berta-Thompson}, {Blecic}, {Casewell}, {Crossfield}, {Crouzet}, {Cubillos},
  {Decin}, {D{\'e}sert}, {Feinstein}, {Gibson}, {Harrington}, {Heng},
  {Henning}, {Kempton}, {Krick}, {Lagage}, {Lendl}, {Lothringer}, {Mansfield},
  {Mayne}, {Mikal-Evans}, {Palle}, {Schlawin}, {Shorttle}, {Wheatley}, \&
  {Yurchenko}}]{tsai:2023}
{Tsai}, S.-M., {Lee}, E. K.~H., {Powell}, D., {et~al.} 2023, \nat, 617, 483

\bibitem[{{Virtanen} {et~al.}(2020){Virtanen}, {Gommers}, {Oliphant},
  {Haberland}, {Reddy}, {Cournapeau}, {Burovski}, {Peterson}, {Weckesser},
  {Bright}, {van der Walt}, {Brett}, {Wilson}, {Millman}, {Mayorov}, {Nelson},
  {Jones}, {Kern}, {Larson}, {Carey}, {Polat}, {Feng}, {Moore}, {VanderPlas},
  {Laxalde}, {Perktold}, {Cimrman}, {Henriksen}, {Quintero}, {Harris},
  {Archibald}, {Ribeiro}, {Pedregosa}, {van Mulbregt}, \& {SciPy 1. 0
  Contributors}}]{virtanen:2020}
{Virtanen}, P., {Gommers}, R., {Oliphant}, T.~E., {et~al.} 2020, Nature
  Methods, 17, 261

\bibitem[{{Wallack} {et~al.}(2024){Wallack}, {Batalha}, {Alderson},
  {Scarsdale}, {Adams Redai}, {Aguichine}, {Alam}, {Gao}, {Wolfgang},
  {Batalha}, {Kirk}, {L{\'o}pez-Morales}, {Moran}, {Teske}, {Wakeford}, \&
  {Wogan}}]{wallack:2024}
{Wallack}, N.~L., {Batalha}, N.~E., {Alderson}, L., {et~al.} 2024, arXiv
  e-prints, arXiv:2404.01264

\bibitem[{{Weiss} {et~al.}(2018){Weiss}, {Marcy}, {Petigura}, {Fulton},
  {Howard}, {Winn}, {Isaacson}, {Morton}, {Hirsch}, {Sinukoff}, {Cumming},
  {Hebb}, \& {Cargile}}]{weiss:2018}
{Weiss}, L.~M., {Marcy}, G.~W., {Petigura}, E.~A., {et~al.} 2018, \aj, 155, 48

\bibitem[{{Welbanks} {et~al.}(2024){Welbanks}, {Bell}, {Beatty}, {Line},
  {Ohno}, {Fortney}, {Schlawin}, {Greene}, {Rauscher}, {McGill}, {Murphy},
  {Parmentier}, {Tang}, {Edelman}, {Mukherjee}, {Wiser}, {Lagage}, {Dyrek}, \&
  {Arnold}}]{welbanks:2024}
{Welbanks}, L., {Bell}, T.~J., {Beatty}, T.~G., {et~al.} 2024, \nat, 630, 836

\bibitem[{{Winn}(2010)}]{winn:2010}
{Winn}, J.~N. 2010, ArXiv e-prints, arXiv:1001.2010

\bibitem[{{Winn} \& {Fabrycky}(2015)}]{winn:2015}
{Winn}, J.~N., \& {Fabrycky}, D.~C. 2015, \araa, 53, 409

\bibitem[{{Wizinowich} {et~al.}(2000){Wizinowich}, {Acton}, {Shelton},
  {Stomski}, {Gathright}, {Ho}, {Lupton}, {Tsubota}, {Lai}, \&
  {Max}}]{wizinowich:2000}
{Wizinowich}, P., {Acton}, D.~S., {Shelton}, C., {et~al.} 2000, \pasp, 112, 315

\bibitem[{{Wolfgang} {et~al.}(2016){Wolfgang}, {Rogers}, \&
  {Ford}}]{wolfgang:2016}
{Wolfgang}, A., {Rogers}, L.~A., \& {Ford}, E.~B. 2016, \apj, 825, 19

\bibitem[{{Yu} {et~al.}(2018){Yu}, {Huber}, {Bedding}, \& {Stello}}]{yu:2018rv}
{Yu}, J., {Huber}, D., {Bedding}, T.~R., \& {Stello}, D. 2018, \mnras, 480, L48

\bibitem[{{Zechmeister} {et~al.}(2018){Zechmeister}, {Reiners}, {Amado},
  {Azzaro}, {Bauer}, {B{\'e}jar}, {Caballero}, {Guenther}, {Hagen}, {Jeffers},
  {Kaminski}, {K{\"u}rster}, {Launhardt}, {Montes}, {Morales}, {Quirrenbach},
  {Reffert}, {Ribas}, {Seifert}, {Tal-Or}, \& {Wolthoff}}]{zechmeister:2018}
{Zechmeister}, M., {Reiners}, A., {Amado}, P.~J., {et~al.} 2018, \aap, 609, A12

\end{thebibliography}

\appendix

\section{Ground-based lightcurve follow-up}
\label{sec:sg1}
The \textit{TESS} pixel scale is $\sim 21\arcsec$ pixel$^{-1}$ and photometric apertures typically extend out to roughly 1 arcminute, generally causing multiple stars to blend in the \textit{TESS} aperture. To rule out a nearby eclipsing binary (NEB) as the potential source of a \textit{TESS} detection and attempt to detect the signal on-target, we observed our target stars and the nearby fields as part of the \textit{TESS} Follow-up Observing Program\footnote{https://tess.mit.edu/followup/} Sub Group 1 \citep[TFOP;][]{collins:2019}. In some cases, we also observed in multiple bands across the optical spectrum to check for wavelength dependent transit depth differences, which can be suggestive of a planet candidate false positive. We used the {\tt TESS Transit Finder}, which is a customized version of the {\tt Tapir} software package \citep{Jensen:2013}, to schedule our transit observations. All observations were reduced  using {\tt AstroImageJ} \citep{collins:2017}, except for MuSCAT and MuSCAT2 as noted in Table \ref{tab:transitfollowupsurvey}.

All of our light curve follow-up observations are summarized in Table \ref{tab:transitfollowupsurvey} and all light curve data are available on the {\tt EXOFOP-TESS} website\footnote{\href{https://exofop.ipac.caltech.edu/tess/}{https://exofop.ipac.caltech.edu/tess/}}. We also provide a short summary of each light curve result and an overall final photometric follow-up disposition in the table. For the targets in this survey, we assign four light curve follow-up dispositions (PC, CPC, VPC, VPC+) to indicate differing levels of confidence that a \textit{TESS} detection is on-target, as described below.

The planet candidate (PC) disposition indicates that we either have no light curve follow-up observations, or the light curve observations are unable to confirm that the \textit{TESS}-detected event is on-target relative to Gaia DR3 and TIC version 8 stars. TOI-1716.01, TOI-1744.01, TOI-1174.01, TOI-1777.01, TOI-4638.01, TOI-5726.01, and TOI-6054.01 have a PC disposition and are not listed in Table \ref{tab:transitfollowupsurvey}.

The cleared planet candidate (CPC) disposition indicates that we have confined the \textit{TESS}-detection to occur on the target star relative to all Gaia DR3 and TIC version 8 stars. Using ground-based photometry, we check all stars out to $2\farcm5$ from the target star that are bright enough, assuming a 100\% eclipse, in \textit{TESS}-band to produce the \textit{TESS}-detected depth at mid-transit. To account for possible delta-magnitude differences between \textit{TESS}-band and the follow-up band, and to account for \textit{TESS}-band magnitude errors, we included an extra 0.5 magnitudes fainter in \textit{TESS}-band. For these cases, the transit depth is generally too shallow to detect on-target in ground-based follow-up observations, so we often saturate the target star on the detector to enable a complete search of all necessary fainter nearby stars. Since the \textit{TESS} point-spread-function has full-width-half-maximum of roughly $40\arcsec$, and the irregularly shaped SPOC photometric apertures and circular QLP photometric apertures generally extend to $\sim1\arcmin$ from the target star, we check for events in stars out to $2\farcm5$ from the target star. For a star to be considered ``cleared'' of an NEB signal, we require its light curve to have a flat model residual RMS value  at least a factor of 3 smaller than the eclipse depth required to produce the \textit{TESS} detection in the star. We ensure that the predicted ephemeris uncertainty is covered by at least $\pm3\sigma$ relative to the most precise SPOC or QLP ephemeris available at the time of publication. Finally, we check all nearby star light curves by eye to ensure that no obvious eclipse-like event is present. By process of elimination, we conclude that when all necessary nearby stars are ``cleared'' of NEBs, the transit is indeed occurring on-target, or in a star so close to the target star, that it was not detected by Gaia DR3 and is not in TIC version 8.

The verified planet candidate (VPC) disposition indicates that we have confirmed using ground-based follow-up light curve photometry that the \textit{TESS}-detected event is occurring on-target. This is accomplished using follow-up photometric apertures that are small enough to exclude most or all of the flux from the nearest Gaia DR3 and/or TIC version 8 star that is bright enough to be capable of producing  the \textit{TESS} signal.

The verified planet candidate plus (VPC+) disposition is the same as VPC, except we have measured transit depths in the target star follow-up photometric apertures across several optical bands. We elevate the disposition to VPC+ if no strong ($>3\sigma$) transit depth difference is detected across the bands.

\section{Ground-based high-resolution imaging follow-up}
\label{sec:sg3}

Below we briefly describe our high-resolution imaging observations of
these systems.  These observations (summarized in Table~\ref{tab:hri})
provided the imaging data used as input for the \texttt{TRICERATOPS}
statistical validation analysis described in
Sec.~\ref{sec:validation}.  In all cases, the final processed data
products, stacked images, and sensitivity curves are available from
at the ExoFOP website\footnote{\url{https://exofop.ipac.caltech.edu/}}.

\begin{deluxetable}{l l l l l l l l } 
\tabletypesize{\scriptsize}
\tablecaption{High-resolution imaging observations \label{tab:hri}}
\tablewidth{0pt}
\tablehead{
\colhead{TOI}	&\colhead{Instrument}	&\colhead{Observatory}	&\colhead{PI}	&\colhead{Date}	&\colhead{Filter}	&\colhead{pixel scale ['']}	&\colhead{PSF FWHM ['']}  \\
}
\startdata
1174	&Alopeke	&Gemini-N	&Howell	&2020-02-17	&832nm	&0.01	&0.02  \\
1180	&NESSI	&WIYN	&Everett	&2022-04-21	&832nm	&0.02	&0.06  \\
1184	&NIRI	&Gemini-S	&Matthews	&2019-11-02	&Br-$\gamma$	&0.022	&0.113  \\
1244	&NIRC2	&Keck II	&Gonzales	&2022-06-23	&K	&0.0099	&0.051  \\
1247	&NIRC2	&Keck II	&Gonzales	&2020-05-28	&Br-$\gamma$	&0.0099	&0.052  \\
1301	&Alopeke	&Gemini-N	&Howell	&2020-06-06	&832nm	&0.01	&0.02  \\
1467	&NIRC2	&Keck II	&Gonzales	&2020-09-09	&K	&0.0099	&0.05  \\
1630	&Alopeke	&Gemini-N	&Howell	&2021-10-17	&832nm	&0.01	&0.02  \\
1659	&Alopeke	&Gemini-N	&Crossfield	&2020-06-06	&832nm	&0.01	&0.02  \\
1691	&Alopeke	&Gemini-N	&Crossfield	&2020-06-08	&832nm	&0.01	&0.02  \\
1716	&Alopeke	&Gemini-N	&Howell	&2021-10-24	&832nm	&0.01	&0.02  \\
1723	&ShARCS	&Shane	&Giacalone	&2020-12-01	&Ks	&0.033	&0.209  \\
1739	&Alopeke	&Gemini-N	&Crossfield	&2020-06-08	&832nm	&0.01	&0.02  \\
1744	&NIRC2	&Keck II	&Gonzales	&2020-09-09	&Br-$\gamma$	&0.0099	&0.048  \\
1758	&Alopeke	&Gemini-N	&Howell	&2021-10-17	&832nm	&0.01	&0.02  \\
1768	&PHARO	&Hale	&Ciardi	&2022-02-13	&Br-$\gamma$	&0.025	&0.099  \\
1772	&PHARO	&Hale	&Ciardi	&2021-02-23	&Br-$\gamma$	&0.025	&0.093  \\
1777	&PHARO	&Hale	&Ciardi	&2020-12-04	&Br-$\gamma$	&0.025	&0.146  \\
1782	&Alopeke	&Gemini-N	&Crossfield	&2020-06-06	&832nm	&0.01	&0.02  \\
1798	&Alopeke	&Gemini-N	&Crossfield	&2020-06-06	&832nm	&0.01	&0.02  \\
1799	&PHARO	&Hale	&Ciardi	&2021-02-23	&Br-$\gamma$	&0.025	&0.093  \\
2211	&Alopeke	&Gemini-N	&Howell	&2021-06-27	&832nm	&0.01	&0.02  \\
4638	&NESSI	&WIYN	&Everett	&2022-09-15	&832nm	&0.02	&0.06  \\
5726	&NESSI	&WIYN	&Everett	&2024-02-18	&832nm	&0.02	&0.06  \\
5799	&PHARO	&Hale	&Ciardi	&2023-07-01	&Br-$\gamma$	&0.025	&0.103  \\
6054	&PHARO	&Hale	&Ciardi	&2023-11-29	&Kcont	&0.025	&0.096  \\
\enddata
\end{deluxetable}

\subsection{`Alopeke}
We observed many of our target systems with the `Alopeke dual-channel
speckle imaging instrument on Gemini-North with a pixel scale of
0.01'' and a typical full width at half maximum (FWHM) resolution of
0.02''. `Alopeke provided simultaneous speckle imaging at 562 and 832
nm, but we used only the 832~nm data in our \texttt{TRICERATOPS}
analysis. Typically, several sets of 1,000 short (60~ms) exposures were
taken for each target, then processed with the speckle pipeline of
\citep{howell:2011}.

\subsection{NESSI}
Three stars, TOI-1180 (UT 2022 Apr. 21), TOI-4638 (UT 2022 Sept. 15)
and TOI-5726 (UT 2024 Feb. 18), were observed at the WIYN 3.5~m
telescope on Kitt Peak, Arizona using the speckle imager NESSI
\citep{scott:2018}.  NESSI produces simultaneous speckle images in two
filters, although only the images using a filter centered at
$\lambda_c=832$~~nm were used in our validation.  Each observation
consisted of a set of either 7 or 9 1000-frame 40~ms exposures.  The
field-of-view was confined using a readout region of $256\times256$
pixels for $4.6\times4.6$~arcseconds and speckle data were used to
constrain the presence of additional sources out to a radius of
1.2~arcseconds from each target star.  Additional observations of a
single star near each science target were obtained for calibration of
the underlying PSF.  To reduce the data, we used the pipeline
described by \cite{howell:2011}.  Pipeline products include a
reconstructed image of the field around each target and, measured from
that, a contrast curve representing the relative magnitude limit as a
function of radial separation from the star.  No companion sources
were detected for any of the NESSI targets.

\subsection{ShARCS}
We observed TOI-1723 on 2020 December 2 using the ShARCS camera on the
Shane 3-meter telescope at Lick Observatory
\citep{kupke:2012,gavel:2014,mcgurk:2014}. Observations were taken with the Shane
adaptive optics (AO) system in natural guide star mode in order to search
for nearby, unresolved stellar companions. We collected a single
sequence of observations using a $K_s$ filter ($\lambda_0 = 2.150$
$\mu$m, $\Delta \lambda = 0.320$ $\mu$m), achieving a contrast of 3.50
magnitudes at 0.5 arcseconds and 4.75 magnitudes at 1
arcsecond. Information about our observing strategy can be found in
\citep{savel:2020}. We reduced the data using the publicly
available \texttt{SImMER} pipeline
\citep{savel:2022}.\footnote{https://github.com/arjunsavel/SImMER}
We found no stellar companions within our detection limits. Further
information about these data can be found in Dressing et al. (in
prep).

\subsection{PHARO}
We observed several targets with high-resolution near-infrared AO
imaging at Palomar Observatory with the PHARO instrument
\citep{hayward:2001} behind the P3K natural guide star system
\citep{dekany:2013}. These observations were taken using a standard
5-point quincunx dither pattern with steps of 5'' in either the
narrow-band 2.1686\,$\mu$m Br-$\gamma$ filter or the broader Kcont
filter. PHARO has a pixel scale of 0.025'' and a total field of view
of $\sim$25''. Data were processed as
described by \citep{furlan:2017}.

\subsection{NIRC2}
Several targets were also observed at Keck Observatory with the NIRC2
instrument \citep{wizinowich:2000} on the 10 m Keck II telescope
\citep[see also][]{schlieder:2021}. The infrared imaging was performed
in natural-guide star AO mode with short integration times designed to
obtain maximum depth without saturating on the primary host star,
using either the K-band filter or the narrow-band Brackett-$\gamma$
filter (as an analogue for K-band for brighter stars). We used the
narrow-camera mode with a pixel scale of $\sim$0.01'', and observed in
a three-point dither pattern to avoid the noisy fourth quadrant of the
detector. Data were processed as described by \citep{furlan:2017} and
\cite{schlieder:2021}.

\subsection{NIRI}
We obtained AO imaging of TOI-1184 with the Gemini
Near-Infrared Imager \citep[NIRI;][]{hodapp:2003}. We collected 9
science frames, each with an exposure time of 4.4 s, and dithered the
telescope by $\sim$2'' between each frame to use the science frames as
sky background frames. The target was observed in the 2.166\,$\mu$m
Br-$\gamma$ ﬁlter with a pixel scale of $\sim$0.022''. Data were
processed as described by \citep{furlan:2017}.

\newpage
\startlongtable
\setlength{\tabcolsep}{3pt}
\begin{deluxetable}{l c l c c l c}
\tablecaption{Ground-based light curve observations \label{tab:transitfollowupsurvey}}
\tablehead{
\colhead{Observatory} &
\colhead{Ap (m)} &
\colhead{Location} &
\colhead{UTC Date} &
\colhead{Filter} &
\colhead{Result} &
\colhead{Disp.$^{a}$}
}
\startdata
TOI-1180.01 & & & & & & \\
\cline{1-1}
GMU$^{b}$              & 0.8  & Fairfax, VA, USA     & 2019-11-10 & R               & no obvious NEBs in 11 NEB check stars & \\
Kotizarovci Obs  & 0.3  & Viskovo, Croatia     & 2020-02-14 & Baader R$^{c}$        & $\sim$1.7 ppt egress in 7.2$\arcsec$ target aperture & \\
Whitin Obs       & 0.7  & Wellesley, MA, USA     & 2020-02-14 & $r'$                  & out-of-transit coverage & \\
MuSCAT3$^{d}$-Hal      & 2.0  & Maui, Hawai'i, USA   & 2022-03-11  & $g',r', i', z_s^{e}$ & $\sim$1.7 ppt transit in 5.3$\arcsec$ target apertures & \\
LCOGT-McD        & 1.0  & Fort Davis, TX, USA  & 2022-05-18  & $B, z_s$             & $\sim$1.7  ppt transit in 5.9$\arcsec$ target apertures & \\
MuSCAT3-Hal       & 2.0  & Maui, Hawai'i, USA  & 2022-04-22  & $g',r', i', z_s$     & $\sim$1.7  ppt transit in 7.2$\arcsec$ target apertures & \multirow{1}{*}{VPC+}\\
[1.5mm]
TOI-1184.01 & & & & & & \\
\cline{1-1}
MuSCAT$^{f}$-NAOJ      & 1.9  & Okayama, Japan & 2020-03-05  & $g',r', z_s$     & $\sim$1 ppt egress in 7.2$\arcsec$ target apertures & \\
MuSCAT2$^{g}$-TCS      & 1.5  & Tenerife, Spain & 2020-05-07  & $g',r', i', z_s$ & possible $\sim$2 ppt 11.7$\arcsec$ target apertures & \\
LCOGT-Hal        & 2.0  & Maui, Hawai'i, USA       & 2020-06-11  & $z_s$            & $\sim$1 ppt transit in 3.6$\arcsec$ target aperture & \\
LCOGT-McD        & 1.0  & Fort Davis, TX, USA      & 2020-08-25  & $z_s$            & $\sim$1 ppt transit in 6.6$\arcsec$ target aperture & \\
LCOGT-Teid       & 1.0  & Tenerife, Canary Islands & 2021-08-07  & $i'$             & $\sim$1 ppt transit in 3.1$\arcsec$ target aperture & \\
LCOGT-Teid       & 1.0  & Tenerife, Canary Islands & 2022-06-22  & $z_s$            & $\sim$1 ppt transit in 4.3$\arcsec$ target aperture & \multirow{1}{*}{VPC}\\
[1.5mm]
TOI-1244.01 & & & & & & \\
\cline{1-1}
GMU              & 0.8  & Fairfax, VA, USA         & 2020-04-13  & R                & $\sim$1 ppt transit in 9.0$\arcsec$ target aperture & \\
MuSCAT3-Hal      & 2.0  & Maui, Hawai'i, USA       & 2021-05-22  & $i'$             & $\sim$1 ppt transit in 4.0$\arcsec$ target aperture & \\
MuSCAT3-Hal      & 2.0  & Maui, Hawai'i, USA       & 2022-07-12  & $g',r', i', z_s$ & $\sim$1 ppt ingress in 4.5$\arcsec$ target apertures & \\
MuSCAT3-Hal      & 2.0  & Maui, Hawai'i, USA       & 2022-08-26  & $g',r', i', z_s$ & $\sim$1 ppt transit in 5.4$\arcsec$ target apertures & \multirow{1}{*}{VPC+}\\
[1.5mm]
TOI-1247.01 & & & & & & \\
\cline{1-1}
LCOGT-McD        & 1.0  & Fort Davis, TX, USA      & 2022-04-05  & $z_s$  &  cleared all NEB check stars & \\
LCOGT-Teid       & 1.0  & Tenerife, Canary Islands & 2022-05-07  & $z_s$  &  cleared 5/6 NEB check stars & \multirow{1}{*}{CPC}\\
[1.5mm]
TOI-1301.01 & & & & & & \\
\cline{1-1}
FLWO$^{h}$       & 1.2  & Amado, Arizona           & 2021-05-22  & $i'$  & $\sim$1 ppt egress in 4.0$\arcsec$ target aperture & \\
LCOGT-Teid       & 1.0  & Tenerife, Canary Islands & 2021-09-08  & $i'$  & $\sim$1 ppt transit in 4.7$\arcsec$ target aperture & \multirow{1}{*}{VPC}\\
[1.5mm]
TOI-1467.01 & & & & & & \\
\cline{1-1}
LCOGT-Teid         & 1.0  & Tenerife, Canary Islands  & 2021-09-14 & $z_s$               & $\sim$1.4 ppt transit in 5.9$\arcsec$ target aperture & \\
MuSCAT3-Hal        & 2.0  & Maui, Hawai'i, USA        & 2022-08-20  & $g',r', i', z_s$ & $\sim$1.4 ppt transit in 4.5$\arcsec$ target apertures & \multirow{1}{*}{VPC+}\\
[1.5mm]
TOI-1630.01 & & & & & & \\
\cline{1-1}
Grand-Pra Obs  & 0.4  & Sion Valais, Switzerland     & 2020-09-30  & $i'$  & cleared 35/36 NEB check stars & \\
LCOGT-McD            & 1.0  & Fort Davis, TX, USA & 2020-11-06  & $z_s$ & $\sim$0.8 ppt transit in 6.2$\arcsec$ target aperture & \\
LCOGT-Teid           & 1.0  & Tenerife, Canary Islands & 2021-11-14  & $z_s$ & $\sim$0.8 ppt transit in 6.6$\arcsec$ target aperture & \multirow{1}{*}{VPC}\\
[1.5mm]
TOI-1659.01 & & & & & & \\
\cline{1-1}
LCOGT-Teid       & 1.0  & Tenerife, Canary Islands    & 2022-10-03   & $z_s$  & cleared all NEB check stars & \multirow{1}{*}{CPC}\\
[1.5mm]
TOI-1691.01 & & & & &  \\
\cline{1-1}
Whitin Obs       & 0.7  & Wellesley, MA, USA  & 2020-02-21   & $r'$ & no obvious NEB & \multirow{1}{*}{PC}\\
[1.5mm]
TOI-1723.01 & & & & & & \\
\cline{1-1}
LCOGT-Teid       & 1.0  & Tenerife, Canary Islands & 2022-11-26 & $z_s$ & cleared all NEBcheck stars  & \multirow{1}{*}{CPC}\\
[1.5mm]
TOI-1739.01 & & & & & & \\
\cline{1-1}
LCOGT-McD        & 1.0  & Fort Davis, TX, USA       & 2022-05-15  & $r'$ & $\sim$1 ppt transit in 8.2$\arcsec$ target aperture & \multirow{1}{*}{VPC}\\
[1.5mm]
TOI-1758.01 & & & & & & \\
\cline{1-1}
LCOGT-McD        & 1.0  & Fort Davis, TX, USA       & 2020-11-05  & B         & $\sim$2 ppt transit in 4.7$\arcsec$ target aperture & \\
LCOGT-Teid       & 1.0  & Tenerife, Canary Islands  & 2022-10-10  & $i'$      & $\sim$2 ppt ingress in 4.7$\arcsec$ target aperture & \\
Kotizarovci Obs  & 0.3  & Viskovo, Croatia          & 2022-10-30  & Baader R &  $\sim$2 ppt ingress in 8.4$\arcsec$ target aperture & \multirow{1}{*}{VPC}\\
[1.5mm]
TOI-1768.01 & & & & & & \\
\cline{1-1}
MuSCAT2-TCS      & 1.5  & Tenerife, Spain          & 2021-01-30  & $g',r', i', z_s$ & cleared 3 of 13 NEBcheck stars & \\
FLWO             & 1.2  & Amado, Arizona           & 2021-02-07  & $i'$             & cleared 10 of 13 NEBcheck stars & \multirow{1}{*}{PC}\\
[1.5mm]
TOI-1772.01 & & & & & & \\
\cline{1-1}
LCOGT-McD        & 1.0  & Fort Davis, TX, USA      & 2022-12-21  & $z_s$ & cleared 5 of 5 NEBcheck stars & \multirow{1}{*}{CPC}\\
[1.5mm]
TOI-1782.03 & & & & & & \\
\cline{1-1}
FLWO             & 1.2  & Amado, Arizona           & 2022-05-19  & $i'$  & $\sim$1.6 ppt transit in 6.2$\arcsec$ target aperture & \\
LCOGT-McD        & 1.0  & Fort Davis, TX, USA      & 2022-05-19  & $i'$  & $\sim$1.6 ppt transit in 6.7$\arcsec$ target aperture & \\
LCOGT-McD        & 1.0  & Fort Davis, TX, USA      & 2023-03-15  & $i'$  & $\sim$1.6 ppt transit in 6.2$\arcsec$ target aperture & \\
LCOGT-McD        & 1.0  & Fort Davis, TX, USA      & 2024-04-12  & $i'$  & $\sim$1.6 ppt ingress in 6.2$\arcsec$ target aperture & \multirow{1}{*}{VPC}\\
[1.5mm]
TOI-1798.01 & & & & & & \\
\cline{1-1}
FLWO             & 1.2  & Amado, Arizona           & 2021-02-09  & $i'$  & no obvious NEBs in 5 NEBcheck stars & \\
LCOGT-Hal        & 0.4  & Maui, Hawai'i, USA       & 2021-04-06  & $i'$  & cleared 5 of 5 NEBcheck stars & \multirow{1}{*}{CPC}\\
[1.5mm]
TOI-1799.01 & & & & &  & \\
\cline{1-1}
LCOGT-McD        & 0.4  & Fort Davis, TX, USA      & 2021-01-02   & $i'$ & cleared 3 of 3 NEBcheck stars & \multirow{1}{*}{CPC}\\
& & & & & & \\
TOI-2211.01 & & & & & & \\
\cline{1-1}
LCOGT-SSO  & 1.0  & Coonabarabran, Australia  & 2021-06-15  & $z_s$ & no obvious NEBs & \multirow{1}{*}{PC}\\
[1.5mm]
TOI-5799.01 & & & & & & \\
\cline{1-1}
TRAPPIST-North   & 0.6   & Marrakech, Morocco        & 2022-09-28  & $z'$             & cleared 213 of 234 NEBcheck stars& \\
LCOGT-Teid       & 1.0   & Tenerife, Canary Islands  & 2023-06-14  & $i'$             & $\sim$2.6 ppt transit in 3.9$\arcsec$ target aperture & \\
MuSCAT2-TCS      & 1.5   & Tenerife, Spain           & 2023-07-08  & $g',r', i', z_s$ & $\sim$2.6 ppt transit in 10.8$\arcsec$ target aperture &\\
TUG$^{i}$        & 1.0   & Antalya, Turkey           & 2023-08-23  & $g'$             &  $\sim$2.6 ppt transit in 5.5$\arcsec$ target aperture & \multirow{1}{*}{VPC+}\\
\cline{1-1}
\enddata
\tablenotetext{a}{The overall follow-up disposition. PC = planetary candidate, CPC $=$ cleared of NEBs, VPC $=$ on-target relative to Gaia DR3 stars, VPC+ $=$ achromatic on-target relative to Gaia DR3 stars. See the text for full disposition definitions.}
\tablenotetext{b} {George Mason University (GMU)}
\tablenotetext{c} {Baader R 610\,nm}
\tablenotetext{d}{Las Cumbres Observatory Global Telescope \citep[LCOGT;][]{brown:2013} 0.4\,m, 1.0\,m, 2.0\,m network nodes at Teide Observatory on the island of Tenerife (TEID), Siding Spring Observatory (SSO), McDonald Observatory (McD), and MuSCAT3 \citep{narita:2020} on Faulkes Telescope North at Haleakala Observatory. Images calibrated by {\tt BANZAI} pipeline \citep{mccully:2018} and photometry extracted using {\tt AstroImageJ} \citep{collins:2017}.}
\tablenotetext{e} {Pan-STARRS $z$-short band ($\lambda_{\rm c} = 8700$\,\AA, ${\rm Width} =1040$\,\AA)}
\tablenotetext{f}{MuSCAT \citep[Multicolor Simultaneous Camera for studying Atmospheres of Transiting exoplanets;][]{narita:2015a} multi-color imager is installed at the 1.88\,m telescope of the National Astronomical Observatory of Japan (NAOJ) in Okayama, Japan. MuSCAT data were extracted using the custom pipeline described in \citep{fukui:2011}.}
\tablenotetext{g}{MuSCAT2 \citep{narita:2019} 4-color multi-band simultaneous camera on the 1.52{\~{}}m Telescopio Carlos S$\backslash$'anchez (TCS). Data reduction follows \citet{parviainen:2020}.}
\tablenotetext{h}{Fred L. Whipple Observatory / KeplerCam}
\tablenotetext{i}{TUBITAK National Observatory (TUG)}
\end{deluxetable}

\end{document}